# A Conditional Model of Wind Power Forecast Errors and Its Application in Scenario Generation

Zhiwen Wang, Chen Shen*, Feng Liu

Dep. of Electrical Engineering, Tsinghua University, Beijing, 100084, China

**ABSTRACT**

In power system operation, characterizing the stochastic nature of wind power is an important albeit challenging issue. It is well known that distributions of wind power forecast errors often exhibit significant variability with respect to different forecast values. Therefore, appropriate probabilistic models that can provide accurate information for conditional forecast error distributions are of great need. On the basis of Gaussian mixture model, this paper constructs analytical conditional distributions of forecast errors for multiple wind farms with respect to forecast values. The accuracy of the proposed probabilistic models is verified by using historical data. Thereafter, a fast sampling method is proposed to generate scenarios from the conditional distributions which are non-Gaussian and interdependent. The efficiency of the proposed sampling method is verified.

**HIGHLIGHTS**

Conditional distributions of forecast errors for multiple wind farms under different forecast values.

A fast scenario generation method for non-Gaussian interdependent distributions.

# 1. Introduction

Nowadays, a large amount of wind power has been integrated into power systems. In power system operation, a wind power forecasting tool plays an important role. Since the forecast values always deviate from the true ones more or less, the resulting forecast errors should be taken into account in generation scheduling [1]. In industrial practice, systems operators usually allocate reserves to compensate the forecast errors [2]. On one hand, if the forecast errors are overestimated, reserves will be overcommitted, increasing operation costs; on the other hand, if the forecast errors are underestimated, reserves will be undercommitted, causing wind spillage and load shedding. Therefore, modeling the wind power forecast errors is a crucial issue for unit commitment (UC) and economical dispatch (ED).

Given an effective point forecasting tool, distributions of wind power forecast errors are conditioned on forecast values. In the literature [3-7], various probabilistic distributions have been adopted to model conditional distributions of wind power forecast errors. In [3], the authors point out that forecast errors of a single wind farm are far from Gaussian distributions, as the kurtosis could be over 10 ( 3 for the Gaussian). Beta distribution is suggested to model forecast error uncertainties. In a relevant study [4], the authors combine the Beta distribution and Dirac delta function, and obtain a "mixed beta distribution", improving the model accuracy. Further, Bruninx *et al* find that Beta distribution is not able to fully characterize the skewed and heavy-tailed forecast errors [5]. To solve this problem, the Levy $\alpha$-stable distribution is adopted. The test results in [5] show that the Levy $\alpha$-stable distribution outperforms Beta distribution. As the distributions of forecast errors are quite various under different forecast levels, the "versatile distribution" with three adjustable parameters is proposed in [6], achieving higher flexibility. Because the "versatile distribution" has more adjustable parameters than Beta/Gaussian distributions, it can better represent forecast error uncertainties. Following a similar idea, Menemenlis *et al* use the time-varying Gamma-like distribution, whose parameters are adjusted as functions of forecast levels, to model the forecast errors [7]. These detailed conditional models [3-7] help the generation scheduling dynamically adjust reserves to different forecast levels. However, they are applicable to the single wind farm case only. They cannot handle multivariate random variables.

To model a joint distribution for adjacent wind farms, the Copula technique has drawn much attention lately. In [8], the Gaussian Copula is used to model the spatial interdependence structure in forecast uncertainties for multiple wind farms across a region. In a similar study [9], applying the Gaussian Copula theory, the authors conduct a multivariate probabilistic analysis for spatial correlated wind generation in the European grid. A remarkable advantage of using the Copula theory to model forecast error uncertainties is made in [10]. Different types of Copula functions, e.g., *Gaussian*, *t*, *Clayton*, *Frank*, *Gumbel*, are adopted to model the stochastic dependence of uncertainties. The Copula-based conditional distributions of forecast errors for multiple wind farms are obtained. For applications, authors in [11] propose a Copula-based chance-constrained optimization model for power system planning. Further, in order to deal with different dependency structures between pairs of random variables, e.g., wind and solar, the vine-Copula methods are investigated in [12], [13], improving the accuracy in high-dimension cases. Although constructing the Copula-based conditional distributions for multiple wind farms has been investigated in the literature [8-13], it is hard to ensure that the constructed distributions have some desirable attributes. For instance, in terms of the scenario generation[1], the Gaussian Copula method generates original scenarios from a Gaussian distribution, transforms the original scenarios into the Copula domain, and obtains final scenarios by using inverse transformations of marginal cumulative distribution functions (CDF). The procedure is time-consuming relative to sampling directly from a joint distribution. The scenario generation procedures of other types of Copulas are more complicated.

In order to incorporate forecast error uncertainties into UC and ED, scenarios generated from the conditional distributions of multiple wind farms are needed [14], [15]. Generally, when random variables are non-Gaussian and interdependent, generating scenarios, i.e., *sampling*, from their joint distribution is difficult [16]. Many existing techniques are either not efficient enough or less accurate [17], [18]. For example, the acceptance-rejection method and conditional sampling method need many steps and multiple transformations, which are time-consuming. The affine transformation method does not ensure that the generated scenarios strictly follow the predefined joint distribution, which may lead to inaccurate results. The Nataf technique is used in [19] to produce wind power scenarios. A time-varying correlation matrix is used in [20] for generating short-term wind uncertainty scenarios. Neither of them proves that the generated scenarios follow a predefined joint distribution. Using historical time series data of wind power and the kernel density estimator, Xydas *et al* propose a generation method for forecast scenarios [21]. Alternatively, Morales *et al* adopt the autoregressive model to generate time series data for wind power scenarios [22]. However, these techniques do not retain the original distributions of uncertainties [12]. To the best of the authors' knowledge, when the conditional forecast error distribution of multiple wind farms is available, there is not an accurate and efficient sampling method that can generate scenarios from the non-Gaussian and interdependent joint distribution.

To address these important issues, this paper aims at a systematic methodology that can accurately model conditional

[1] A scenario can be understood as a plausible realization of uncertainty [14]. The uncertainty could be formulated as random variables or a stochastic process. Wind power scenario generation means producing a set of possible realizations of wind power uncertainty. From the prospective of the probability theory, the "scenario generation" indeed means generating samples from a given probabilistic distribution. In this sense, we abuse the terminology "*sampling*" to stand for "scenario generation" in this paper. A rigorous definition and several illustrative examples can be found in [14].

distributions and generate scenarios. The original contributions are twofold:

(1) On the basis of Gaussian mixture model (GMM), this paper constructs conditional distributions of wind power forecast errors for multiple wind farms under different forecast values. With the proposed distributions, non-Gaussianity and correlations of forecast error uncertainties can be handled. What's more, operator can conveniently obtain the conditional distribution of the aggregated forecast errors across a region.

(2) Based on the proposed probabilistic model, a fast method is developed to generate scenarios for wind power forecast errors with high accuracy and efficiency. The method is proved to be an exactly accurate sampling method for interdependent random variables. With the proposed sampling method, tens of thousands of scenarios can be generated within milliseconds.

The rest of this paper is organized as follows. In Section 2, a framework for the proposed methodology is provided. In Section 3, the GMM is used to represent a joint distribution of actual wind power outputs and forecast values for multiple wind farms. In Section 4, analytical formulae are derived to construct a conditional distribution of wind power forecast errors. Advantages of the proposed probabilistic model are discussed. In Section 5, a method that generates scenarios from the constructed conditional distribution is proposed. Case study results are presented in Section 6. Conclusions and limitations are shown in Section 7.

## 2. Framework

The proposed methodology consists of three phases. A flow chart is shown in Fig. 1:

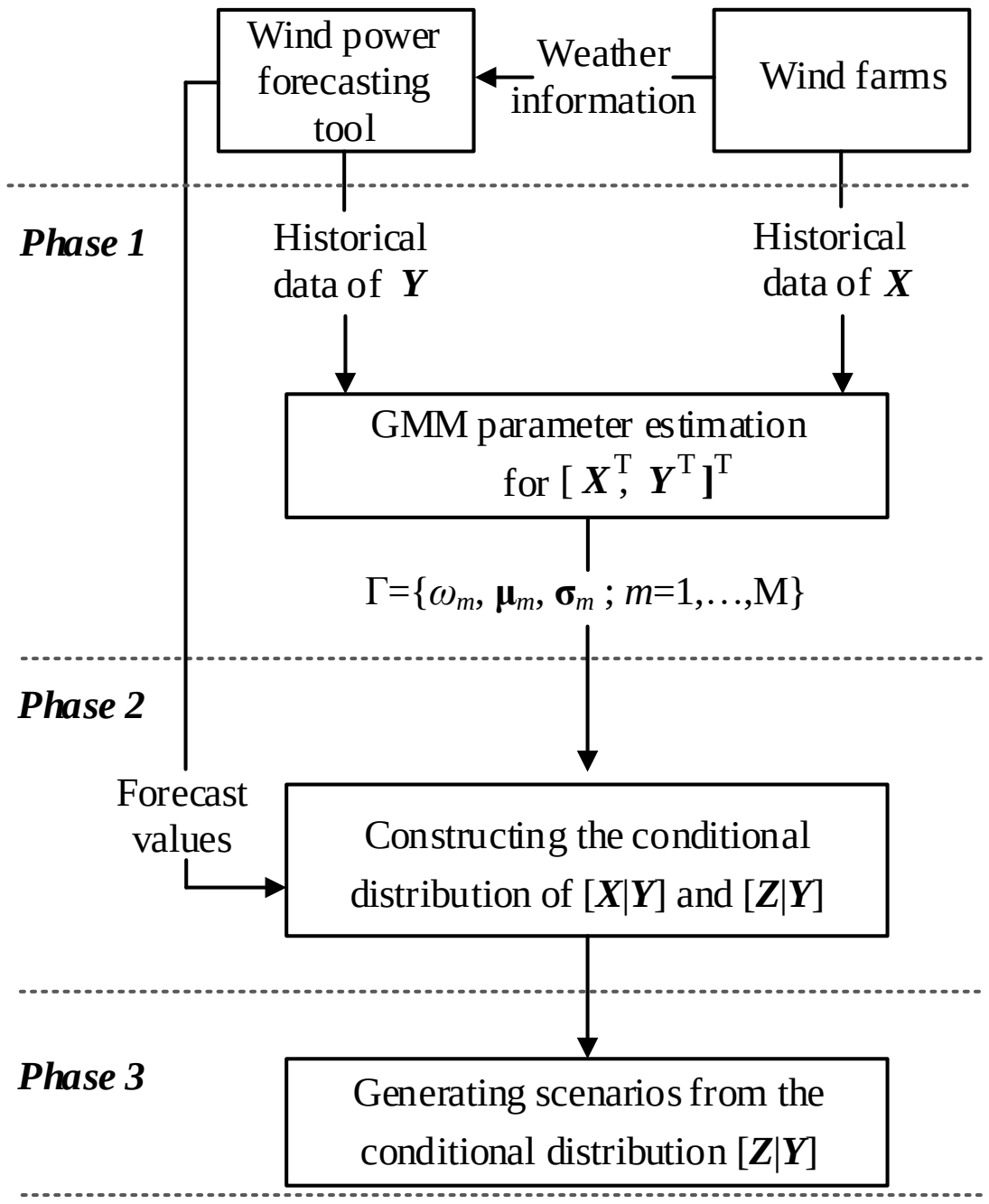

**Fig. 1** Implementation procedure of the proposed methodology.

***Phase 1:*** Modeling a probability density functions (PDF) of the actual wind power outputs and forecast values by a GMM. Without loss of generality, let a random vector $\boldsymbol{X}$ denote actual wind power outputs of multiple wind farms, $\boldsymbol{Y}$ denote the corresponding forecast values, and $\boldsymbol{Z}$ denote the forecast errors. A GMM is used to represent the joint PDF of an aggregated random vector $[\boldsymbol{X}^{\mathrm{T}}\ \boldsymbol{Y}^{\mathrm{T}}]^{\mathrm{T}}$.

***Phase 2:*** Constructing conditional distributions of wind power with respect to forecast values $[\boldsymbol{X}\,|\,\boldsymbol{Y}]$, as well as the conditional distributions of wind power forecast errors $[\boldsymbol{Z}\,|\,\boldsymbol{Y}]$.

***Phase 3:*** Generating scenarios from the constructed conditional distributions of $[\boldsymbol{Z}\,|\,\boldsymbol{Y}]$.

In the implementation procedure, the GMM parameter set Γ of $[\boldsymbol{X}^{\mathrm{T}}\ \boldsymbol{Y}^{\mathrm{T}}]^{\mathrm{T}}$ is estimated according to historical data in the first phase. That is, the parameter set Γ of a GMM is obtained off-line. The parameter set Γ only needs to be updated when there is a need to update the historical dataset, e.g. once a day.

## 3. Joint distribution of actual power outputs and forecast values

### *3.1 Date source*

The hourly wind power data used in this paper is from the National Renewable Energy Laboratory (NREL) public dataset "*eastern wind integration data set*". The data set consists of wind resource and plant output data for the eastern United States. The meteorological data, e.g., wind speed, were generated on the basis of two meteorological models: the Mesoscale Atmospheric Simulation System and the Weather Research and Forecasting model. Then, the wind power outputs were produced using turbine power curves of IEC Class 1 and 2. The forecast values were produced by running a statistical point forecasting tool called *SynForecast*. There are three forecasting lead time horizons in the dataset: next-day, 6-hour, and 4-hour. NREL has compared the simulated data (wind power outputs and forecasts) with real measurements. The results show that the simulated data and the measurements are very similar. That is to say, although the wind power outputs and their forecasts were simulated, they were verified to well represent the stochastic nature of real-world wind power uncertainties. A detailed description of the dataset is available in [23] and [24]. It should be noted that the NREL wind power data records do not have outliers. When other raw data sources are used, the outliers should be carefully preprocessed. Two alternative approaches are presented in [25] and [26].

The dates of the NREL data range from 20040101 to 20061231. There are 26304 data records. In this paper, the data of 2004 (8784 records) is used as a training set for modeling uncertainty, while the data of 2005 (8760records) is used as a test set for the scenario generation test.

### *3.2 Wind power uncertainty*

Usually, forecasting tools have forecast errors. That is, the forecast values deviate from the actual wind power outputs. Given a point forecasting tool, the distribution of forecast errors varies significantly with respect to different forecast values. To clearly illustrate the phenomenon, this paper uses the historical data of a wind farm in Illinois and generates the histograms of the forecast errors under different forecasts. The ID of this wind farm in the NREL dataset is 4209. The installed capacity is 1014 MW. The maximum values of actual wind power and forecast values are 984MW, 983MW, respectively. The maximum forecast error is 654MW. The data is hourly. The forecast lead time is 4 hour. The 8784 data records in 2004 are used. The histograms of the forecast errors with respect to different forecast values are obtained as follows:

(1) The historical data pairs of actual wind power and forecasts $[\boldsymbol{X}^{\mathrm{T}}\ \boldsymbol{Y}^{\mathrm{T}}]^{\mathrm{T}}$ are normalized to the installed capacity.

(2) The historical data pairs of $[\boldsymbol{X}^{\mathrm{T}}\ \boldsymbol{Y}^{\mathrm{T}}]^{\mathrm{T}}$ are transformed into wind power forecast errors and forecasts $[\boldsymbol{Z}^{\mathrm{T}}\ \boldsymbol{Y}^{\mathrm{T}}]^{\mathrm{T}}$.

(3) The historical data pairs of $[\boldsymbol{Z}^{\mathrm{T}}\ \boldsymbol{Y}^{\mathrm{T}}]^{\mathrm{T}}$ are categorized into several bins on the basis of the forecasts. A bin consists of a central value $y^*$ and a width $wd$. The bin ranges from $y^*-wd$ to $y^*+wd$. In this paper, the number of bins is 9. It can be changed to other values, if needed. The value of $y^*$ is set to be 0.1 through 0.9. The width of each bin is 0.05. For example, the first bin is [0.05, 0.15], the second bin is [0.15, 0.25], and the last bin is [0.85, 0.95].

(4) The data pairs, whose forecast values are within the bin $[y^*-wd\ ,\ y^*+wd]$, constitute the histograms of $\boldsymbol{Z}$ conditioned on $y^*$.

For demonstration purpose, the histograms of the 1st, 5th, and 9th bins are shown in Fig.2, while the others are omitted. It can be observed that the distributions of forecast errors conditioned on $y^*$ are quite different. In [3], [27], Beta distributions are used to approximate those histograms one by one, while the authors in [4], [5], and [6] report that the accuracy of Beta still needs improvement.

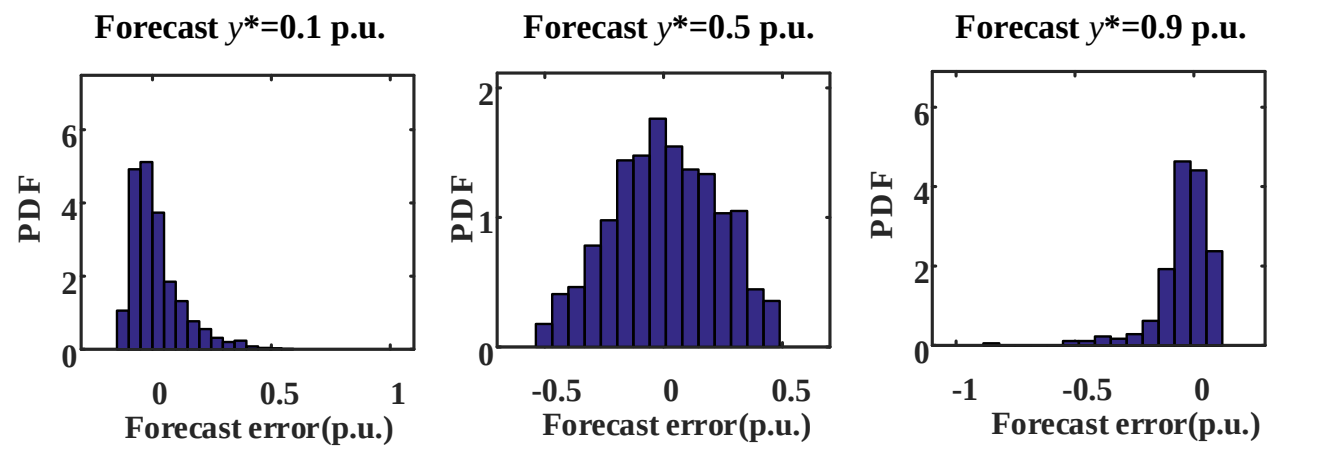

**Fig. 2** Histograms of forecast errors of the 1st, 5th, and 9th bins. Others are omitted. The forecast lead time is 4 hour. The data is from a single wind farm (ID: 4209) in the NREL "eastern wind integration data set". The data is normalized to the installed capacity.

Beyond the variability, wind power uncertainties of adjacent wind farms have stochastic dependence. In generation scheduling, there are many cases wherein the outputs and forecast errors of multiple wind farms are needed. For example, when transmission limits are considered, one needs to know the output of each wind farm. So far, the only practical option for modeling a multidimensional random vector has been limited to a multivariate Gaussian distribution. However, the Gaussian assumption of wind uncertainty is not accurate [5], [6]. Modeling the dependence of non-Gaussian interdependent random variables (actual wind power outputs, forecast values, and forecast errors) with acceptable accuracy remains challenging.

To handle the variability and dependence of forecast errors appropriately, this paper models forecast errors in the following sequence: first, a GMM is adopted to represent a joint distribution of actual wind power outputs and forecast values. Then, a conditional distribution of forecast errors is constructed analytically.

Some factors contribute to conditional distributions of forecast errors, e.g., the forecast lead time, the technique that a forecasting tool utilizes, and wind farm locations. This paper does not discuss how those factors affect the probabilistic distributions of the forecast errors, nor is it aimed at developing a new probabilistic forecasting tool. Rather, this paper is focused on modeling conditional distributions of forecast errors, utilizing historical data of a given point forecasting tool.

This paper uses the GMM to model forecast errors. The GMM parameters are estimated directly from the historical data of $[\boldsymbol{X}^{\mathrm{T}}\ \boldsymbol{Y}^{\mathrm{T}}]^{\mathrm{T}}$. When the forecasting tool and the lead time change, the historical data records of $\boldsymbol{Y}$ change accordingly. In such a situation, the GMM parameters should be estimated again using the new historical data of $[\boldsymbol{X}^{\mathrm{T}}\ \boldsymbol{Y}^{\mathrm{T}}]^{\mathrm{T}}$. In other words, different forecasting tools and lead times correspond to different historical data records of $[\boldsymbol{X}^{\mathrm{T}}\ \boldsymbol{Y}^{\mathrm{T}}]^{\mathrm{T}}$, which lead to different GMM parameters. For a given forecasting tool with a certain forecast lead time, as long as the historical data of $[\boldsymbol{X}^{\mathrm{T}}\ \boldsymbol{Y}^{\mathrm{T}}]^{\mathrm{T}}$ is available, the proposed method can be used to model forecast errors. Therefore, the two factors, i.e., the forecasting tool and the lead time, do not limit the practicality of the proposed method.

As far as wind farm location is concerned, the parameters of the GMMs for different wind farms might be different depending on the historical data. For example, the parameters of a GMM for wind farm A are estimated using historical data of A, while the parameters for wind farm B are estimated using historical data of B. Since the historical data of A and B are different, the estimated parameters of GMMs are different. As a result, the conditional distributions of A and B are different.

### *3.2 Gaussian mixture model*

In data clustering and machine learning, the GMM is known for its high level accuracy in characterizing multiple random variables [28]. Recently, several researchers have applied the GMM technique to power system uncertainty analysis and verified its superiority in modeling stochastic power outputs of renewable energy and loads [29], [30].

A GMM for an aggregated random vector, $[\boldsymbol{X}^{\mathrm{T}}\ \boldsymbol{Y}^{\mathrm{T}}]^{\mathrm{T}}$, is defined as a convex combination of multivariate Gaussian distribution functions with an adjustable parameter set $\Gamma=\{\omega_m, \boldsymbol{\mu}_m, \boldsymbol{\sigma}_m\ ;\ m=1,\ldots,\mathrm{M}\}$. A mathematical expression of the GMM is given as follows:

$$f_{\boldsymbol{XY}}\left(\boldsymbol{x},\boldsymbol{y}\right)=\sum_{m=1}^{\mathrm{M}}\omega_m N_m\left(\boldsymbol{x},\boldsymbol{y}\ ;\ \boldsymbol{\mu}_m,\boldsymbol{\sigma}_m\right) \tag{1}$$

$$\sum_{m=1}^{\mathrm{M}}\omega_m=1,\quad \omega_m>0 \tag{2}$$

$$N_m\left(\boldsymbol{x},\boldsymbol{y}\ ;\ \boldsymbol{\mu}_m,\boldsymbol{\sigma}_m\right)=\frac{e^{-\frac{1}{2}\left(\begin{bmatrix}\boldsymbol{x}\\\boldsymbol{y}\end{bmatrix}-\boldsymbol{\mu}_m\right)^{\mathrm{T}}\boldsymbol{\sigma}_m^{-1}\left(\begin{bmatrix}\boldsymbol{x}\\\boldsymbol{y}\end{bmatrix}-\boldsymbol{\mu}_m\right)}}{\left(2\pi\right)^{\mathrm{W}}\det\left(\boldsymbol{\sigma}_m\right)^{1/2}} \tag{3}$$

$$\boldsymbol{\mu}_m=\begin{bmatrix}\boldsymbol{\mu}_m^{x}\\\boldsymbol{\mu}_m^{y}\end{bmatrix},\boldsymbol{\sigma}_m=\begin{bmatrix}\boldsymbol{\sigma}_m^{xx} & \boldsymbol{\sigma}_m^{xy}\\\boldsymbol{\sigma}_m^{yx} & \boldsymbol{\sigma}_m^{yy}\end{bmatrix} \tag{4}$$

where $f_{XY}(\boldsymbol{x},\boldsymbol{y})$ is the joint PDF of $[\boldsymbol{X}^{\mathrm{T}}\ \boldsymbol{Y}^{\mathrm{T}}]^{\mathrm{T}}$; $\omega_m$ is the weight coefficient; W in the denominator of Eq. (3) denotes the number of wind farms; $N_m(\cdot)$ denotes a multivariate Gaussian distribution function, which is called the $m$th Gaussian component of the GMM; M is the total number of Gaussian components; $\boldsymbol{\mu}$ and $\boldsymbol{\sigma}$ with subscripts and superscripts are parameters.

Determining the parameter set $\Gamma$ of the GMM is a typical parameter estimation problem. With historical data of $\boldsymbol{X}$ and $\boldsymbol{Y}$, one can obtain the parameter set $\Gamma$ using the maximum likelihood estimation (MLE) technique. A well-known algorithm is the Expectation Maximization (EM) algorithm. Many commercial software tools provide reliable off-the-shelf solvers for estimating the parameters of GMM, e.g., *gmdistribution.fit* in MATLAB. Guidelines about the GMM parameter estimation are available in [29].

## 4. Conditional distributions of forecast errors

In theory, a conditional distribution of $\boldsymbol{X}$ can be computed as a joint PDF of $[\boldsymbol{X}^{\mathrm{T}}\ \boldsymbol{Y}^{\mathrm{T}}]^{\mathrm{T}}$ divided by a marginal distribution of $\boldsymbol{Y}$. That is:

$$f_{X|Y}\left(\boldsymbol{x}\mid\boldsymbol{y}\right)=\frac{f_{XY}\left(\boldsymbol{x},\boldsymbol{y}\right)}{f_{Y}\left(\boldsymbol{y}\right)} \tag{5}$$

However, computing the marginal distribution of $\boldsymbol{Y}$ usually requires a multiple integral operation, which cannot be computed easily in practice. As a consequence, constructing the conditional distribution of $\boldsymbol{X}$ from the joint PDF of $[\boldsymbol{X}^{\mathrm{T}}\ \boldsymbol{Y}^{\mathrm{T}}]^{\mathrm{T}}$ is not trivial. To circumvent the problem, it is helpful to use some properties of the GMMs. In the following, the details for constructing the conditional distribution from a GMM are given, followed by discussions of the possible extensions and advantages of the proposed method.

### *4.1 Conditional distribution of a GMM*

***Proposition 1***: if the joint PDF of a random vector $[\boldsymbol{X}^{\mathrm{T}}\ \boldsymbol{Y}^{\mathrm{T}}]^{\mathrm{T}}$ is represented by a GMM in Eq. (1), then the marginal distribution of $\boldsymbol{Y}$ is also a GMM:

$$f_{Y}\left(\boldsymbol{y}\right)=\sum_{m=1}^{\mathrm{M}}\omega_m N_m\left(\boldsymbol{y}\ ;\ \boldsymbol{\mu}_m^{yy},\boldsymbol{\sigma}_m^{yy}\right) \tag{6}$$

***Proof***:

Note that the random vector $\boldsymbol{Y}$ can be regarded as a linear transformation of $[\boldsymbol{X}^{\mathrm{T}}\ \boldsymbol{Y}^{\mathrm{T}}]^{\mathrm{T}}$:

$$\boldsymbol{Y}=\begin{bmatrix}\mathbf{0} & \mathbf{1}\end{bmatrix}\begin{bmatrix}\boldsymbol{X}\\\boldsymbol{Y}\end{bmatrix} \tag{7}$$

On the basis of the so-called "*linear invariance*" property of the GMM [31], the distribution of $\boldsymbol{Y}$ can be computed as follows:

$$\begin{aligned}f_{Y}\left(\boldsymbol{y}\right)&=\sum_{m=1}^{\mathrm{M}}\omega_m\times N_m\left(\boldsymbol{y}\,;\begin{bmatrix}\mathbf{0} & \mathbf{1}\end{bmatrix}\boldsymbol{\mu}_m,\begin{bmatrix}\mathbf{0} & \mathbf{1}\end{bmatrix}\boldsymbol{\sigma}_m\begin{bmatrix}\mathbf{0} & \mathbf{1}\end{bmatrix}^{\mathrm{T}}\right)\\&=\sum_{m=1}^{\mathrm{M}}\omega_m N_m\left(\boldsymbol{y}\ ;\ \boldsymbol{\mu}_m^{yy},\boldsymbol{\sigma}_m^{yy}\right)\end{aligned} \tag{8}$$

***Proposition 2***: if the joint PDF of a random vector $[\boldsymbol{X}^{\mathrm{T}}\ \boldsymbol{Y}^{\mathrm{T}}]^{\mathrm{T}}$ is represented by a GMM in Eq. (1), then the conditional distribution of $\boldsymbol{X}$ with respect to $\boldsymbol{Y}=\boldsymbol{y}$ is also a GMM:

$$f_{X|Y}\left(\boldsymbol{x}\mid\boldsymbol{y}\right)=\sum_{m=1}^{\mathrm{M}}\omega_m' N_m\left(\boldsymbol{x}\mid\boldsymbol{y}\,;\boldsymbol{\mu}_m^{x\bullet y},\boldsymbol{\sigma}_m^{xx\bullet y}\right) \tag{9}$$

where

$$\omega_m'=\omega_m\frac{N_m\left(\boldsymbol{y}\ ;\ \boldsymbol{\mu}_m^{yy},\boldsymbol{\sigma}_m^{yy}\right)}{\sum_{l=1}^{\mathrm{M}}\omega_l N_l\left(\boldsymbol{y}\ ;\ \boldsymbol{\mu}_l^{yy},\boldsymbol{\sigma}_l^{yy}\right)} \tag{10}$$

$$\boldsymbol{\mu}_m^{x\bullet y}=\boldsymbol{\mu}_m^{x}+\boldsymbol{\sigma}_m^{xy}\left(\boldsymbol{\sigma}_m^{yy}\right)^{-1}\left(\boldsymbol{y}-\boldsymbol{\mu}_m^{y}\right) \tag{11}$$

$$\boldsymbol{\sigma}_m^{xx\bullet y}=\boldsymbol{\sigma}_m^{xx}-\boldsymbol{\sigma}_m^{xy}\left(\boldsymbol{\sigma}_m^{yy}\right)^{-1}\boldsymbol{\sigma}_m^{yx} \tag{12}$$

***Proof***:

Substituting Eq. (1) and Eq. (6) into Eq. (5), one can obtain the derivations as follows

$$\begin{aligned}f_{X|Y}\left(\boldsymbol{x}\mid\boldsymbol{y}\right)&=\frac{\sum_{l=1}^{\mathrm{M}}\omega_m N_m\left(\boldsymbol{x},\boldsymbol{y}\ ;\ \boldsymbol{\mu}_m,\boldsymbol{\sigma}_m\right)}{\sum_{l=1}^{\mathrm{M}}\omega_l N_l\left(\boldsymbol{y}\ ;\ \boldsymbol{\mu}_l^{yy},\boldsymbol{\sigma}_l^{yy}\right)}\\&=\sum_{l=1}^{\mathrm{M}}\left[\frac{\omega_m N_m\left(\boldsymbol{y}\ ;\ \boldsymbol{\mu}_m^{yy},\boldsymbol{\sigma}_m^{yy}\right)}{\sum_{l=1}^{\mathrm{M}}\omega_l N_l\left(\boldsymbol{y}\ ;\ \boldsymbol{\mu}_l^{yy},\boldsymbol{\sigma}_l^{yy}\right)}N_m\left(\boldsymbol{x}\mid\boldsymbol{y}\,;\boldsymbol{\mu}_m^{x\bullet y},\boldsymbol{\sigma}_m^{xx\bullet y}\right)\right]\\&=\sum_{m=1}^{\mathrm{M}}\omega_m' N_m\left(\boldsymbol{x}\mid\boldsymbol{y}\,;\boldsymbol{\mu}_m^{x\bullet y},\boldsymbol{\sigma}_m^{xx\bullet y}\right)\end{aligned} \tag{13}$$

In the second step of Eq. (13), an equation of a multivariable Gaussian distribution function is used, which is shown in Eq. (14). The proof of Eq. (14) can be found in [32].

$$\begin{aligned}&N_m\left(\boldsymbol{x},\boldsymbol{y}\ ;\ \boldsymbol{\mu}_m,\boldsymbol{\sigma}_m\right)\\&=N_m\left(\boldsymbol{y}\ ;\ \boldsymbol{\mu}_m^{yy},\boldsymbol{\sigma}_m^{yy}\right)\times N_m\left(\boldsymbol{x}\mid\boldsymbol{y}\,;\boldsymbol{\mu}_m^{x\bullet y},\boldsymbol{\sigma}_m^{xx\bullet y}\right)\end{aligned} \tag{14}$$

Once the joint PDF of $[\boldsymbol{X}^{\mathrm{T}}\ \boldsymbol{Y}^{\mathrm{T}}]^{\mathrm{T}}$ is represented by a GMM, the conditional distribution of $\boldsymbol{X}$ with respect to $\boldsymbol{Y}=\boldsymbol{y}$ is analytically computed using Eqs. (9)-(12)

Note that the forecast error $\boldsymbol{Z}$ can be regarded as a difference between the actual wind power and the corresponding forecast value. Hence, a conditional distribution of forecast errors with respect to forecast values is given as follows:

$$\boldsymbol{Z}=\boldsymbol{X}-\boldsymbol{Y} \tag{15}$$

$$f_{Z|Y}\left(\boldsymbol{z}\mid\boldsymbol{y}\right)=\sum_{l=1}^{\mathrm{M}}\omega_l' N_m\left(\boldsymbol{z}+\boldsymbol{y}\mid\boldsymbol{y};\boldsymbol{\mu}_m^{x\bullet y},\boldsymbol{\sigma}_m^{xx\bullet y}\right) \tag{16}$$

### *4.2 Possible extensions*

Wind power forecasts are often performed several times per day. For example, for the wind power at period $t$, forecasts are issued at $t$-24, $t$-6, and $t$-4. An example is provided in Fig. 3. Let $\boldsymbol{X}_t$ denote the actual wind power at period $t$. Let $\boldsymbol{Y}_{t\text{-}24}$, $\boldsymbol{Y}_{t\text{-}6}$, and $\boldsymbol{Y}_{t\text{-}4}$ denote the forecast values of 24h-ahead, 6h-ahead, and 4h-ahead lead times, respectively. Usually, the forecasts of different lead times ($\boldsymbol{Y}_{t\text{-}24}$, $\boldsymbol{Y}_{t\text{-}6}$, and $\boldsymbol{Y}_{t\text{-}4}$) are correlated. The resulting forecast errors ($\boldsymbol{X}_t-\boldsymbol{Y}_{t\text{-}24}$, $\boldsymbol{X}_t-\boldsymbol{Y}_{t\text{-}6}$, and $\boldsymbol{X}_t-\boldsymbol{Y}_{t\text{-}4}$) are also correlated. In order to model the interdependence of the forecast values/errors, joint distributions of forecast values/errors are needed.

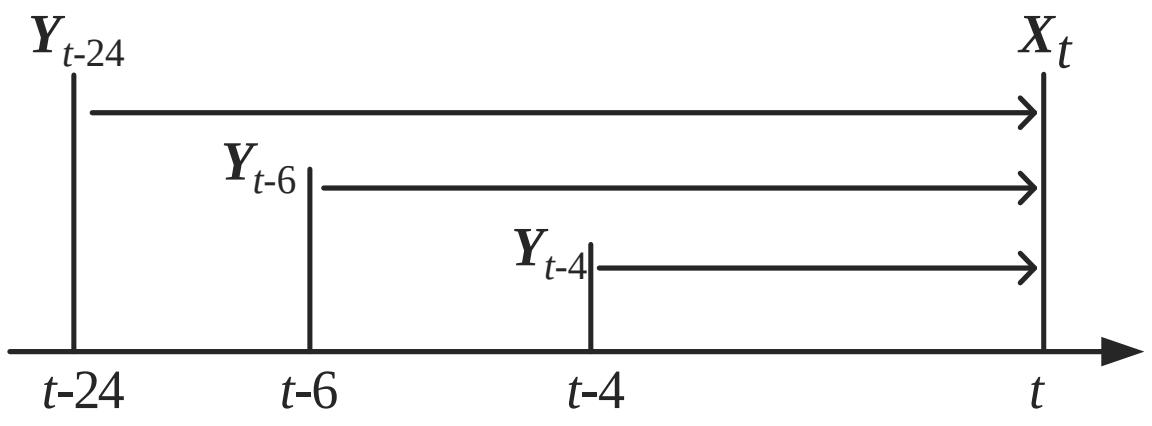

**Fig. 3** Forecast values with different lead times

In this subsection, let $\boldsymbol{Y}$ represent an aggregated random vector $[\boldsymbol{Y}_{t\text{-}24}^{\mathrm{T}}, \boldsymbol{Y}_{t\text{-}6}^{\mathrm{T}}\ \ \boldsymbol{Y}_{t\text{-}4}^{\mathrm{T}}]^{\mathrm{T}}$ and $\boldsymbol{X}$ represent $\boldsymbol{X}_t$. The GMM is used to represent the joint PDF of $[\boldsymbol{X}^{\mathrm{T}}\ \boldsymbol{Y}^{\mathrm{T}}]^{\mathrm{T}}$. For brevity, the joint distribution is still denoted by Eq. (1). It should be noted that the dimension of $\boldsymbol{X}$ remains at W, while the dimension of $\boldsymbol{Y}$ changes from W to 3W. The power of 2π in the denominator of Eq. (3) changes from W to 2W.

### *4.2.1 Joint distribution of forecast values with different lead times*

Note that the joint distribution of $\boldsymbol{Y}$=$[\boldsymbol{Y}_{t\text{-}24}^{\mathrm{T}}, \boldsymbol{Y}_{t\text{-}6}^{\mathrm{T}}\ \ \boldsymbol{Y}_{t\text{-}4}^{\mathrm{T}}]^{\mathrm{T}}$ is the marginal distribution of $[\boldsymbol{X}^{\mathrm{T}}\ \boldsymbol{Y}^{\mathrm{T}}]^{\mathrm{T}}$. On the basis of ***Proposition 1***, the joint distribution of $\boldsymbol{Y}$=$[\boldsymbol{Y}_{t\text{-}24}^{\mathrm{T}}, \boldsymbol{Y}_{t\text{-}6}^{\mathrm{T}}\ \ \boldsymbol{Y}_{t\text{-}4}^{\mathrm{T}}]^{\mathrm{T}}$ can be computed as follows:

$$f_{\boldsymbol{Y}}\left(\boldsymbol{y}\right)=\sum_{m=1}^{\mathrm{M}}\omega_m N_m\left(\boldsymbol{y}\ ;\ \boldsymbol{\mu}_m^{yy},\boldsymbol{\sigma}_m^{yy}\right) \tag{17}$$

### *4.2.2 Joint distribution of forecast errors with different lead times*

The forecast errors of $\boldsymbol{Y}_{t\text{-}24}$, $\boldsymbol{Y}_{t\text{-}6}$, and $\boldsymbol{Y}_{t\text{-}4}$ are defined as follows:

$$\boldsymbol{Z}=\begin{bmatrix}\boldsymbol{X}_t-\boldsymbol{Y}_{t-24}\\ \boldsymbol{X}_t-\boldsymbol{Y}_{t-6}\\ \boldsymbol{X}_t-\boldsymbol{Y}_{t-4}\end{bmatrix} \tag{18}$$

Note that $\boldsymbol{Z}$ is a linear transformation of $[\boldsymbol{X}^{\mathrm{T}}\ \boldsymbol{Y}^{\mathrm{T}}]^{\mathrm{T}}$:

$$\boldsymbol{Z}=\underbrace{\begin{bmatrix}\mathbf{1} & \mathbf{-1} & \mathbf{0} & \mathbf{0}\\ \mathbf{1} & \mathbf{0} & \mathbf{-1} & \mathbf{0}\\ \mathbf{1} & \mathbf{0} & \mathbf{0} & \mathbf{-1}\end{bmatrix}}_{\mathbf{A}}\begin{bmatrix}\boldsymbol{X}\\ \boldsymbol{Y}\end{bmatrix} \tag{19}$$

On the basis of the "*linear invariance*" property of the GMM [31], the joint distribution of forecast errors can be immediately obtained by Eq. (20):

$$f_{\boldsymbol{Z}}\left(\boldsymbol{z}\right)=\sum_{m=1}^{\mathrm{M}}\omega_m N_m\left(\boldsymbol{z}\ ;\ \mathbf{A}\boldsymbol{\mu}_m,\mathbf{A}\boldsymbol{\sigma}_m\mathbf{A}^{\mathrm{T}}\right) \tag{20}$$

### *4.2.3 Gaussian case*

This paper uses GMM to model the non-Gaussian distribution of actual wind power outputs $\boldsymbol{X}$ and forecast values $\boldsymbol{Y}$. Note that when the Gaussian component number of the GMM is set to be M=1, the GMM will naturally degenerate to be a Gaussian distribution. As a result, the GMM is able to handle the case that the random variables $[\boldsymbol{X}^{\mathrm{T}}\ \boldsymbol{Y}^{\mathrm{T}}]^{\mathrm{T}}$ are Gaussian.

## *4.3 Advantages*

In several papers [3-7], a number of distributions are adopted to model conditional forecast errors separately at different forecast levels. Compared with these traditional methods, the proposed method has several advantages:

(1) Because the traditional methods need to estimate parameters of conditional distributions at every forecast level, they have a heavy computational burden when there are multiple wind farms. For example, if there are 3 wind farms and 10 forecast levels for each wind farm, then there are $10^3$ combinations of different forecast levels, which means $10^3$ parameters estimations. In contrast, the proposed method estimates parameters of the joint distribution of $[\boldsymbol{X}^{\mathrm{T}}\ \boldsymbol{Y}^{\mathrm{T}}]^{\mathrm{T}}$ only one time, while constructing conditional distributions over all forecast levels.

(2) For a particular forecast level $\boldsymbol{y}^*$, traditional methods usually collect historical data of forecast errors into a bin if their forecast values are near $\boldsymbol{y}^*$, and extract a conditional distribution from the collected data of the bin. With multiple wind farms, the historical data records near $\boldsymbol{y}^*$ may not be sufficient. For example, according the public dataset from NREL, the number of a 1h-resolution dataset of three wind farms (IDs: 4209, 4208, and 4468) for one year is 8784, while the number of the data records near $\boldsymbol{y}^*$=[0.5 0.5 0.5] with a width 0.05 is only 37. With the limited number of data records, it is not easy for traditional methods to construct a conditional distribution of $\boldsymbol{y}^*$. In contrast, the proposed method utilizes all 8784 historical data records to extract the joint distribution of $[\boldsymbol{X}^{\mathrm{T}}\ \boldsymbol{Y}^{\mathrm{T}}]^{\mathrm{T}}$, and computes the conditional distribution of $\boldsymbol{y}^*$.

(3) It is important that the joint distribution is a GMM, which comes with useful attributes. For example, in additional to the distribution of forecast errors for multiple wind farms, operators also want to know the aggregated forecast errors of these wind farms across a region. Define the aggregated wind power $X_{sum}$ and the aggregated forecast value $Y_{sum}$ as follows:

$$\begin{bmatrix}X_{sum}\\ Y_{sum}\end{bmatrix}=\begin{bmatrix}\mathbf{1} & \mathbf{0}\\ \mathbf{0} & \mathbf{1}\end{bmatrix}\begin{bmatrix}\boldsymbol{X}\\ \boldsymbol{Y}\end{bmatrix} \tag{21}$$

where **1/0** are unit/zero vectors with proper dimensions.

Suppose that $[\boldsymbol{X}^{\mathrm{T}}\ \boldsymbol{Y}^{\mathrm{T}}]^{\mathrm{T}}$ is modeled by Copula. Since entries of $[\boldsymbol{X}^{\mathrm{T}}\ \boldsymbol{Y}^{\mathrm{T}}]^{\mathrm{T}}$ are not independent, the convolution technique does not apply. Hence, computing the distributions of $[X_{sum}\ Y_{sum}]^{\mathrm{T}}$ and $[X_{sum} \mid Y_{sum}]$ is difficult [16]. In contrast, if the distribution of $[\boldsymbol{X}^{\mathrm{T}}\ \boldsymbol{Y}^{\mathrm{T}}]^{\mathrm{T}}$ is a GMM, it can be proved that $[X_{sum}\ \ Y_{sum}]^{\mathrm{T}}$ is also a GMM [32]. Its distribution function is:

$$\begin{aligned}&f_{X_{sum},Y_{sum}}\left(x_{sum}\ ,y_{sum}\right)=\sum_{m=1}^{\mathrm{M}}\omega_m\times\\ &N_m\left(x_{sum}\ ,y_{sum}\ ;\begin{bmatrix}\mathbf{1} & \mathbf{0}\\ \mathbf{0} & \mathbf{1}\end{bmatrix}\boldsymbol{\mu}_m,\begin{bmatrix}\mathbf{1} & \mathbf{0}\\ \mathbf{0} & \mathbf{1}\end{bmatrix}\boldsymbol{\sigma}_m\begin{bmatrix}\mathbf{1} & \mathbf{0}\\ \mathbf{0} & \mathbf{1}\end{bmatrix}\right)\end{aligned} \tag{22}$$

Then, one can use Eqs. (9)-(12) to compute the conditional distribution [$X_{sum}$ | $Y_{sum}$]. Note that the aggregated forecast error $Z_{sum}$ is a difference between $X_{sum}$ and $Y_{sum}$:

$$Z_{sum} = X_{sum} - Y_{sum} \tag{23}$$

Finally, the conditional distribution of $Z_{sum}$ with respect to $Y_{sum}$ can be obtained using Eq. (16).

(4) It is important that the conditional distribution of $\boldsymbol{Z}$ is still a GMM, which facilitates the scenario generation.

# 5. Scenario generation

In this section, a theoretical foundation and sampling procedure of a multivariate Gaussian distribution is briefly introduced, followed by details of the proposed method that sample from the constructed conditional GMM.

## *5.1 Generating scenarios from a Gaussian distribution*

If a random vector $\boldsymbol{Z}$ follows a multivariate Gaussian distribution $N_m(\boldsymbol{\mu}_m,\boldsymbol{\sigma}_m)$, then $\boldsymbol{Z}$ can be converted into an independent Gaussian random vector $\boldsymbol{Z'}$, which can be easily sampled using the following linear transformation [33]:

$$\boldsymbol{Z}' = \boldsymbol{C}^{-1}\left(\boldsymbol{Z} - \boldsymbol{\mu}_m\right) \quad , \quad \boldsymbol{C}\,\boldsymbol{C}^{\mathrm{T}} = \boldsymbol{\sigma}_m \tag{24}$$

Based on this principle, one can generate samples $\boldsymbol{S}_{\boldsymbol{Z}'}$ of $\boldsymbol{Z'}$, and obtain samples $\boldsymbol{S}_{\boldsymbol{Z}}$ of $\boldsymbol{Z}$ by Eq. (24). A standard sampling procedure of a Gaussian distribution is provided in [33].

## *5.2 Generating scenarios from a GMM*

Usually, if a random vector $\boldsymbol{Z}$ does not follow a multivariate Gaussian distribution, then a transformation, such as Eq. (24), cannot guarantee the independence. This is one of the main reasons why sampling from a joint distribution of interdependent random variables is difficult. Such an obstacle also exists for GMMs. To circumvent this issue, a new idea is to sample individually from each Gaussian component of the GMM, and then assemble those samples. The procedure of generating $C_{\text{total}}$ scenarios from a GMM $\Sigma\omega_m N_m(\cdot)$ is detailed in as follows.

***Step 1:*** For each Gaussian component $N_m(\cdot)$ of the GMM, the number of samples to generate is assigned as $\omega_m C_{\text{total}}$.

***Step 2:*** Generating $\omega_m C_{\text{total}}$ samples from the multivariate Gaussian distribution $N_m(\cdot)$. Those samples are denoted by $\boldsymbol{S}_{\boldsymbol{Z}}^{m}$.

***Step 3:*** Collecting all those samples $\boldsymbol{S}_{\boldsymbol{Z}}^{m}$ together results in a new sample set $\boldsymbol{S}_{\boldsymbol{Z}}^{\text{GMM}}$:

$$\boldsymbol{S}_{\boldsymbol{Z}}^{\text{GMM}} = \left\{\boldsymbol{S}_{\boldsymbol{Z}}^{1}\ ,\ldots\ldots,\ \boldsymbol{S}_{\boldsymbol{Z}}^{M}\right\} \tag{25}$$

In Appendix A, it is proved that the sample set $\boldsymbol{S}_{\boldsymbol{Z}}^{\text{GMM}}$ follows the predefined joint distribution $\Sigma\omega_m N_m(\cdot)$. Therefore, the proposed method is an exactly accurate sampling methodology.

In the scenario generation procedure, one only needs to sample from a series of Gaussian distributions. This task can be done using a standard sampling function *mvnrnd* in MATLAB within milliseconds. In this regard, the proposed method is a fast sampling methodology.

## *5.3 Comparison to related research*

### *5.3.1 Comparison to the Copula methods*

In several references [8-13], different Copulas are adopted to model joint distributions. In terms of accuracy, the Copula methods use different dependence structures to achieve a satisfactory fitting performance. The GMM can increase the number of components and adjust the parameter set Γ to improve its accuracy. In this regard, both Copula and GMM are potentially appropriate tools for uncertainty modeling. However, in terms of scenario generation, the GMM has one significant advantage, which is explained as follows.

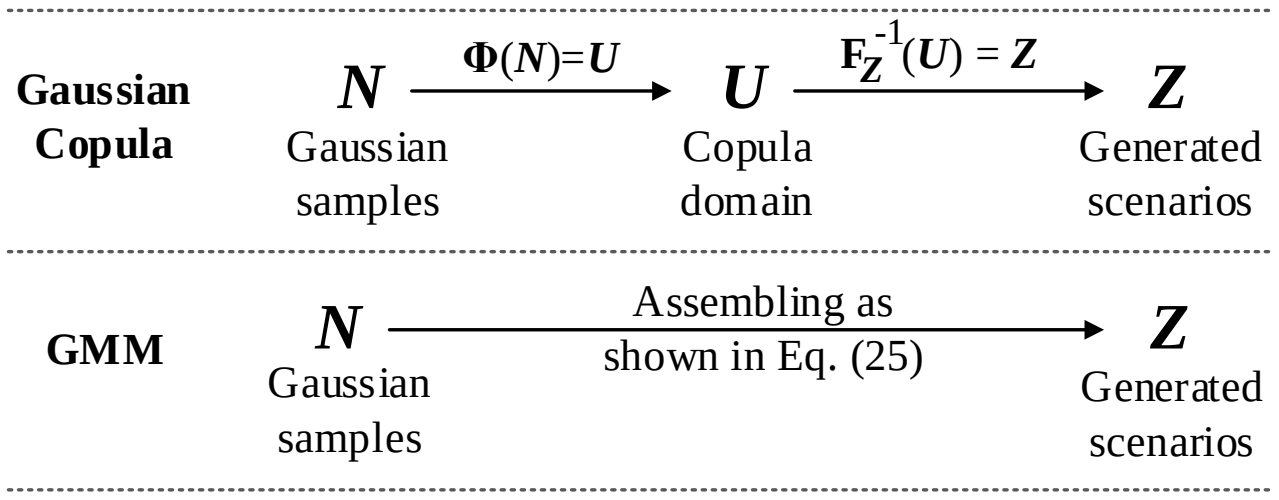

**Fig. 4** Illustration of Gaussian Copula and GMM in scenarios generation

The Gaussian Copula method is used as an example. The sampling procedures of the Gaussian Copula and the GMM methods are illustrated in Fig. 4. The Gaussian Copula transforms original Gaussian samples $\boldsymbol{N}$ to the Copula domain as $\boldsymbol{U}$, and obtains final scenarios $\boldsymbol{Z}$ using the inverse cumulative distribution function (CDF) operation. In this case, one has to construct CDFs for each random variable, and then tautologically find values of inverse CDFs by numerical search techniques. These steps are time-consuming. In contrast, the GMM method directly generates samples in $\boldsymbol{Z}$ domain. No transformation or numerical search is needed. Thus, the proposed GMM method saves computational effort and time.

### *5.3.2 Comparison to existing GMM research*

(1) In [35], Ke *et al* use a customized GMM to represent unconditional distributions of wind power. The conditional distributions of forecast errors are not discussed in [34].

(2) In [31], the authors use the GMM to model $\boldsymbol{X}$, considering the correlation among adjacent wind farms. The conditional distributions of forecast errors are not discussed. This paper models $[\boldsymbol{X}^{\mathrm{T}}\ \boldsymbol{Y}^{\mathrm{T}}]^{\mathrm{T}}$ by a GMM. The proposed method, which is used to construct conditional distributions of [$\boldsymbol{X}$ | $\boldsymbol{Y}$] and [$\boldsymbol{Z}$ | $\boldsymbol{Y}$] from the joint distribution of $[\boldsymbol{X}^{\mathrm{T}}\ \boldsymbol{Y}^{\mathrm{T}}]^{\mathrm{T}}$, is not reported in [31].

(3)The scenario generation method is not reported in [31] or [34].

## 6. Case study

### 6.1 Results of modeling conditional distributions

#### 6.1.1 Single wind farm case

The data of a wind farm (ID: 4209) with a 4-hour forecast lead time horizon is used for this test. The joint distribution of $[\boldsymbol{X}^{\mathrm{T}}\ \boldsymbol{Y}^{\mathrm{T}}]^{\mathrm{T}}$ is modeled by the GMM with 20 components. Thereafter, conditional distributions of forecast errors $\boldsymbol{Z}$ with respect to forecasts $\boldsymbol{Y}= y^*$ from 0.1 to 0.9 are computed. The constructed conditional distributions are compared with the histograms of conditional forecast errors. The way to obtain the histograms conditioned on different $y^*$ is detailed in Section 3.2. Comparative results are shown in Fig. 5. It can be seen that the conditional distributions constructed from the GMM are consistent with the histograms under different forecasts.

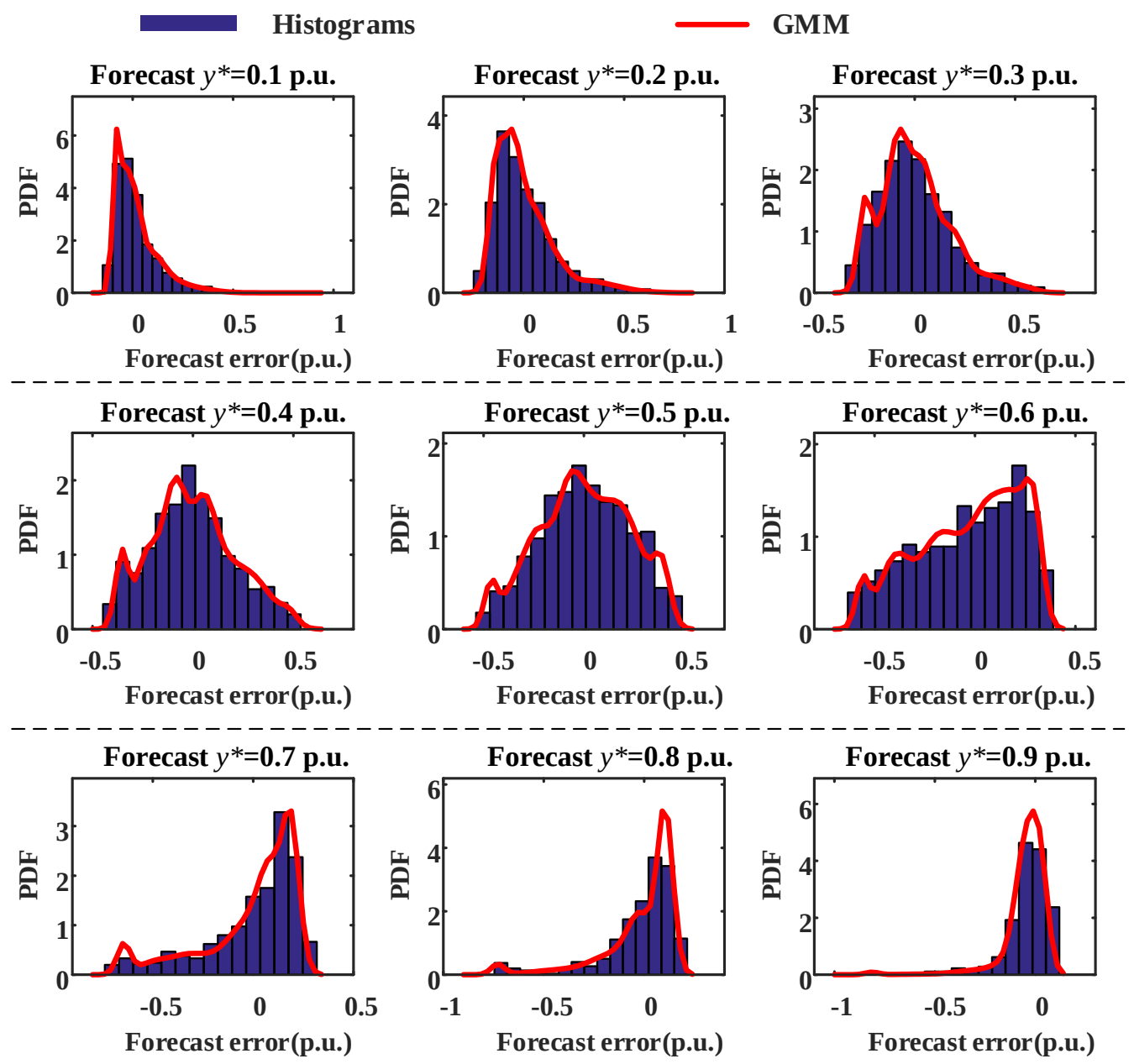

**Fig. 5** Histograms of forecast errors of the 9 bins. The forecast lead time is 4 hour. The data is from a single wind farm (ID: 4209) in the NREL "eastern wind integration data set". The data is normalized to the installed capacity.

To quantify the accuracy of different methods modeling wind power forecast errors, two indices are used. The first one is the log-likelihood function value [35] for evaluating the fitting performance of historical data of $[\boldsymbol{X}^{\mathrm{T}}\ \boldsymbol{Y}^{\mathrm{T}}]^{\mathrm{T}}$. The second one is the root-mean-square error (RMSE) for conditional distributions of $[\boldsymbol{Z} \mid \boldsymbol{Y}]$. The RMSE is defined as follows:

$$\mathrm{RMSE}=\sqrt{\frac{1}{n}\sum_{i=1}^{n}\left(pdf_i^{\mathrm{Cons}}-pdf_i^{\mathrm{His}}\right)^2} \tag{26}$$

where $pdf_i^{\mathrm{Cons}}$ is the PDF of the constructed conditional distribution; $pdf_i^{\mathrm{His}}$ is the PDF of the histogram; $n$ is the total number of points on the PDF curve.

In this test, Gaussian Copula and t Copula are compared with the GMM. There are three reasons why this paper chooses the Gaussian/t Copulas. First, the Gaussian/t Copula methods are the most popular ones in the literature [8-13]. Second, the Gaussian/t Copula methods not only model joint distributions, but also construct conditional distributions. They match well with the scope of this paper. The method using Gaussian/t Copulas to construct conditional distributions is detailed in [10]. Third, the Gaussian/t Copula methods have standard testing functions in commercial software tools, e.g., MATLAB. Hence, it is convenient for readers to reproduce the test results in this paper. The code used in the tests follows standard guidelines from MATLAB documentation on Copulas [36]. Quantitative test results are shown in Fig. 6 and Table 1. It can be seen that:

1) The GMM is better at fitting $[\boldsymbol{X}^{\mathrm{T}}\ \boldsymbol{Y}^{\mathrm{T}}]^{\mathrm{T}}$ than the Gaussian/t Copula methods, as the GMM increases the log-likelihood values by 22%, 18%, respectively.

2) The GMM has an advantage in representing conditional distributions than the Gaussian/t Copula methods, as its RMSEs under every forecast level from 0.1 to 0.9 p.u. are the lowest.

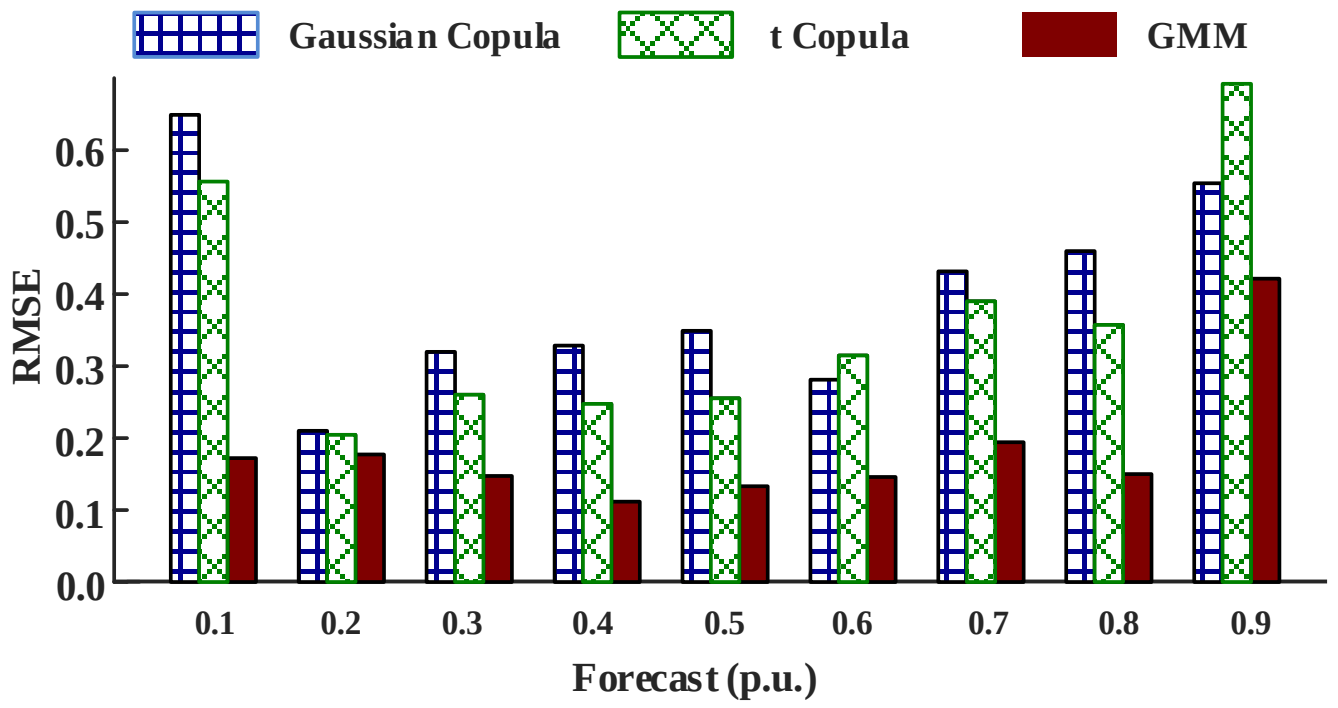

**Fig. 6** Accuracy comparison for modeling conditional distributions

**Table 1**
Fitting test of modeling $[\boldsymbol{X}^{\mathrm{T}}\ \boldsymbol{Y}^{\mathrm{T}}]^{\mathrm{T}}$

| Methods | Log-likelihood function values($10^3$) |
|---|---|
| Gaussian Copula | 5.6905 |
| t Copula | 5.9104 |
| GMM | 6.9614 |

#### 6.1.2 Multiple wind farms case

The data of three wind farms (IDs: 4209, 4208, and 4468) with a 4-hour forecasting lead time horizon is used for this test. Since the dimensions of $[\boldsymbol{X} \mid \boldsymbol{Y}]$ are 3, it is difficult to use a picture like Fig. 5 to visualize the fitting performance. However, if the joint distribution of $[\boldsymbol{X}^{\mathrm{T}}\ \boldsymbol{Y}^{\mathrm{T}}]^{\mathrm{T}}$ is represented by the GMM, it is feasible to analytically compute the conditional distributions of the aggregated forecast errors $Z_{sum}$ with respect to the aggregated forecast values $Y_{sum}$ using the derivations in Section 4.3. On the basis of this idea, this paper uses the aggregation $Z_{sum}|Y_{sum}$ to visualize the fitting performance. Results are shown in Fig. 7. Note that if $[\boldsymbol{X}^{\mathrm{T}}\ \boldsymbol{Y}^{\mathrm{T}}]^{\mathrm{T}}$ is modeled by Copula methods, it is difficult to compute the conditional distributions of the aggregated wind power forecast errors. Hence, results of Copula methods are not provided. From Fig. 7, it can be seen that the conditional distributions of the aggregated forecast errors under different aggregated forecasts fit the histograms well, indicating that the GMM has a satisfactory level of accuracy.

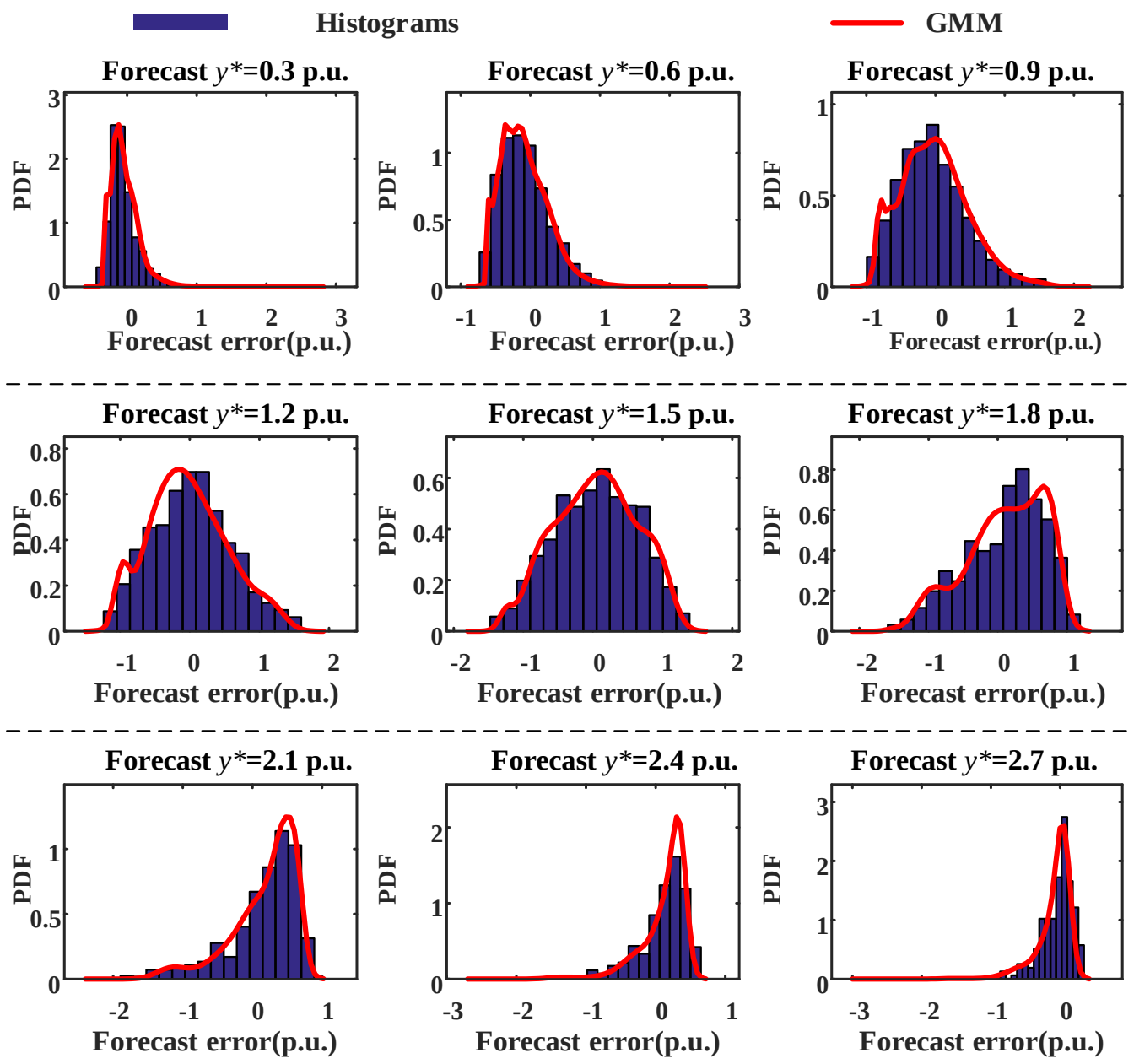

**Fig. 7** Histograms of the aggregated forecast errors. The forecast lead time is 4 hour. The data is from three single wind farms (ID: 4209, 4208, and 4468) in the NREL "eastern wind integration data set". The data is normalized to the installed capacities.

Furthermore, this paper provides an illustrative example in Fig. 8 for the joint distributions of forecast errors on the condition of given forecast values: 0.2 p.u. for WF1, 0.4 p.u. for WF2, and 0.4 p.u. for WF3. It can be seen that conditional forecast errors are interdependent among different wind farms. Results of Fig. 8 are consistent with those reported in [10].

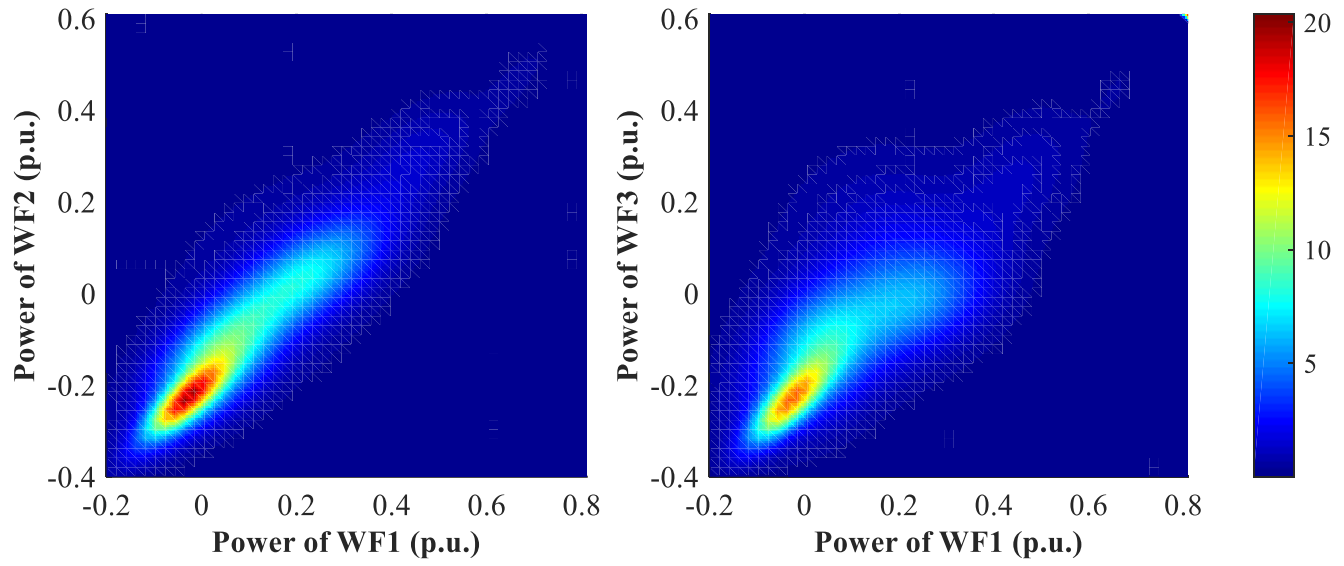

**Fig. 8** .

### 6.1.3 *Tests with real-world data*

In addition to the NREL data, in this paper, the proposed modeling method is also tested with real-world data from Bonneville Power Administration (BPA) [37]. The data is hourly. The forecast lead time is 4 hour. The dates of the BPA data range from 20160101 to 20161231. The maximum actual wind power output is 4493MW. The maximum forecast wind power is 4491MW. The data is rated to 4500MW. The test results are shown in Fig. 9. It can be seen that the GMM-based conditional distributions coincide with the histograms.

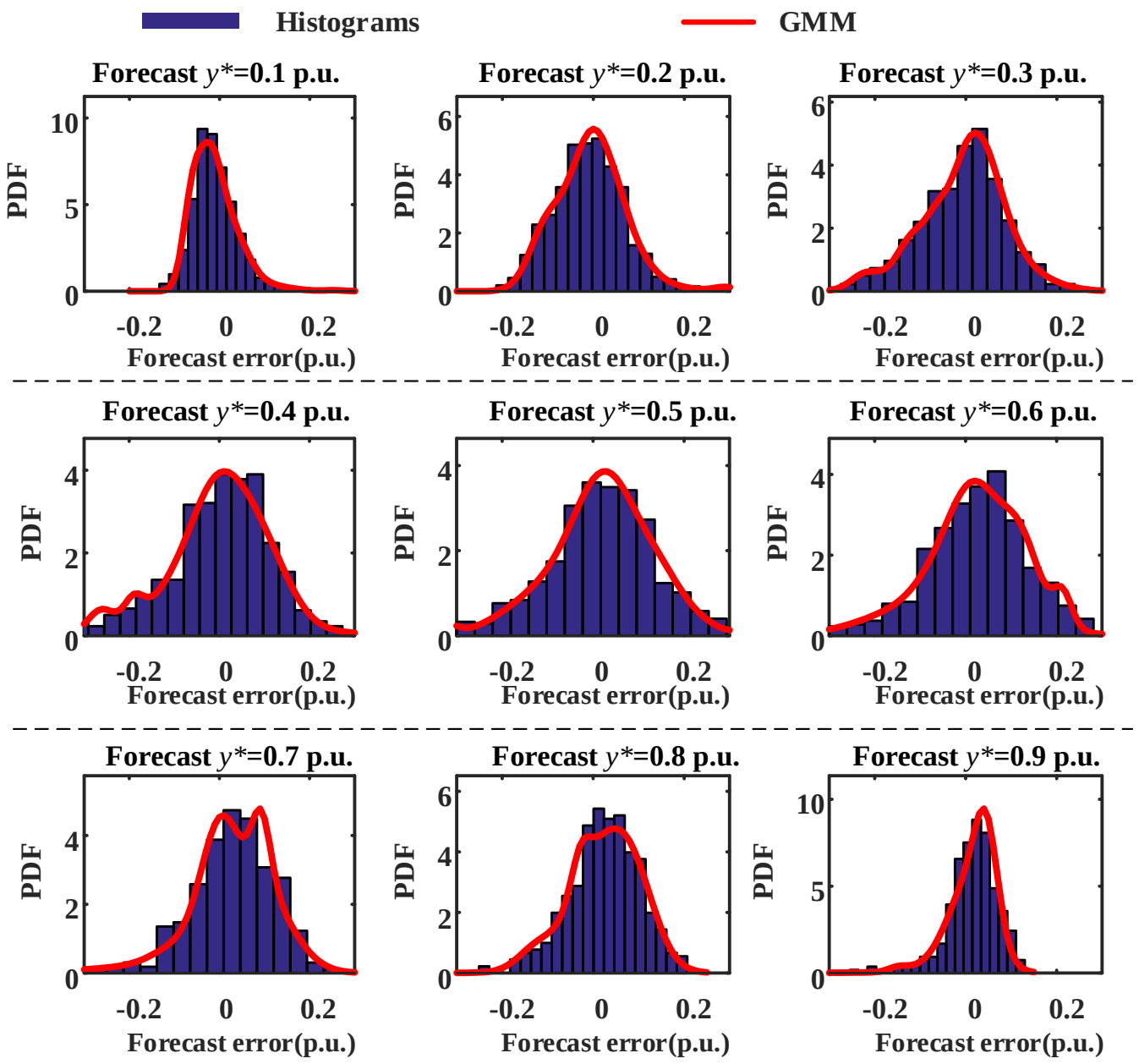

**Fig. 9** Histograms of the BPA forecast errors. The forecast lead time is 4 hour. The data is from BPA. The data is normalized to 4500MW.

### 6.1.4 *Different forecast lead times and locations*

(1) Different forecast lead times

There are test results of different forecast lead times, e.g., 6-hour, the next-day. They are provided in Appendix B. The test results show that the GMM has a satisfactory performance.

(2) Different wind farm locations

The proposed method is tested using the data of four wind farms, which are located in different places: Texas (ID: 81), Oklahoma (ID: 764), Minnesota (ID: 2943), and Iowa (ID: 391), respectively. Their installed capacities are 1118, 1120, 1107, and 1001 MW. The forecast lead time is 4 hour. The data records are normalized to the installed capacities. The test results are provided in Appendix C. According to Fig C1, the GMMs match well with the histograms of the four wind farms. That is to say, the GMM can represent wind forecast error uncertainties in different locations.

### 6.1.5 *Test result of possible extensions*

In Section 4.2, the proposed method is modified to compute the joint distributions of forecasts and the forecast errors with different lead times. The test results of the possible extensions are provided in Appendix D.

## 6.2 *Results of generating scenarios*

In scenario generation, the proposed method is compared with three other methods in the literature. The efficiency and effectiveness of the four methods are tested, respectively.

As far as the efficiency is concerned, different methods are used to generate 1000, 5000, 10000, and 50000 scenarios. The tests are implemented on a Core-i5 2.39-GHz processor. The test results are listed in Table 2. It can be seen that the proposed method greatly improves the sampling efficiency with respect to the Copula methods, especially when the number of scenarios is large. The efficient performance benefits from the GMM-based sampling method that generates scenarios directly in ***Z*** domain. As a comparison, the Copula methods have to transform original scenarios from ***U*** domain to ***Z*** domain by numerically searching inverse CDFs. The transformations are not time-saving.

**Table 2**
Efficiency comparison

| Number of scenarios | Time cost (seconds) | | | |
|---|---|---|---|---|
| | 1000 | 5000 | 10000 | 50000 |
| t Copula method | 3.83 | 19.74 | 41.58 | 217.47 |
| Gaussian Copula method | 3.91 | 18.64 | 38.52 | 203.81 |
| ARMA [35] | 0.0040 | 0.0052 | 0.0066 | 0.0353 |
| The proposed method | 0.0023 | 0.0032 | 0.0042 | 0.0263 |

***Remark 1***: The ARMA method [38] is a popular one used in the literature to generate time series data of wind power uncertainty. However, unlike the GMM and Copula methods, the ARMA method cannot provide probabilistic distributions of forecast errors for operators. Therefore, it does not match the scope of this paper. It is provided in the Table 2 for reference purpose only. What's more, although the ARMA method is more efficient than the Copula methods, it does not outperform the proposed GMM method.

***Remark 2***: in the literature, there are some other methods for wind power scenario generation [20], [39]. Because these methods require to compute the inverse CDFs in the scenario generation procedure, they suffer from the same problem as the Copula methods.

***Remark 3***: If the number of wind farms increases, the time cost will increase, too. A test is conducted to show the time costs of different scenario generation methods with different number of wind farms. According to the results shown in Fig. 10, it takes about 750s for the Gaussian Copula method to generate 50000 scenarios for 10 wind farms. As a comparison, the GMM-based method still costs very little time (0.07232s).

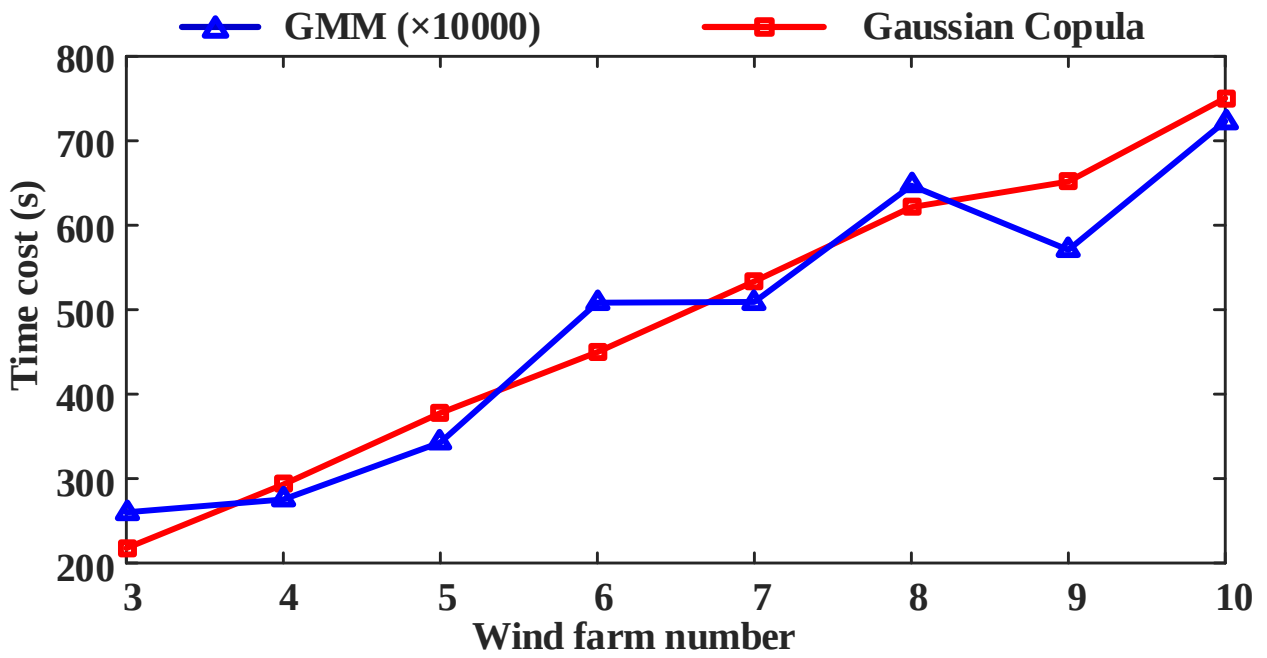

**Fig. 10** Time costs of scenario generation with different wind farm number. The time cost of the GMM method is multiplied by 10000.

When a larger amount of scenarios are generated, one wants to retain $K$ representative scenarios. The *K-means* clustering [40] method is used for scenario reduction in this paper.

In order to evaluate the effectiveness of generated scenarios, a 52-week test is performed. In every week, two indices, namely, *mean absolute error* (MAE) and *variance* (VAR), are used to assess the quality of the generated scenarios:

$$\text{MAE} = \frac{1}{168}\sum_{t=1}^{168}\left|\sum_{k=1}^{K}\pi_k\left(z_k^t + y^t - x^t\right)\right| \tag{27}$$

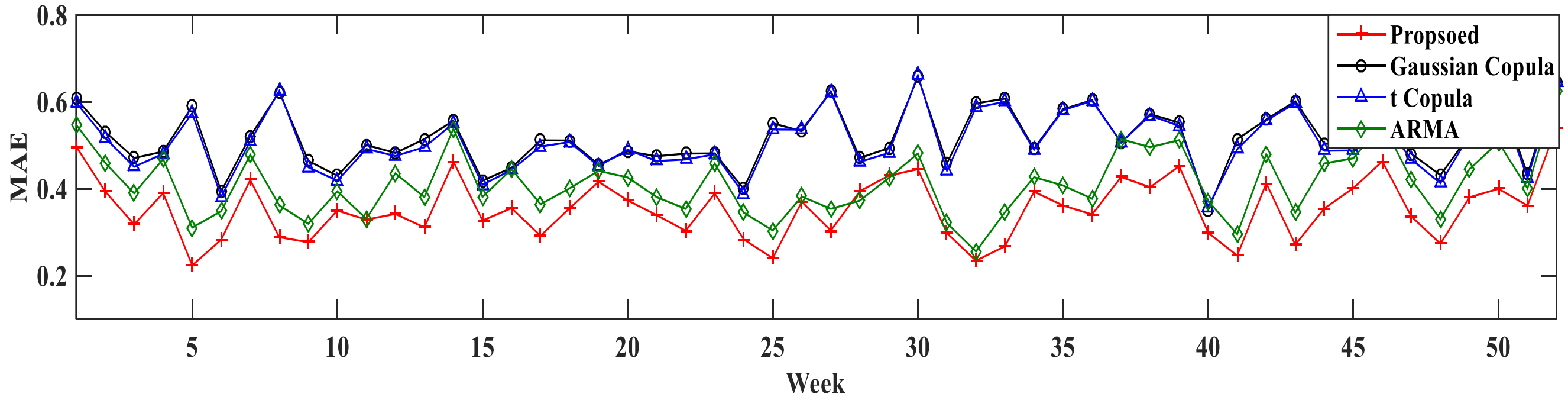

**Fig. 11** MAEs of different methods in 52 weeks.

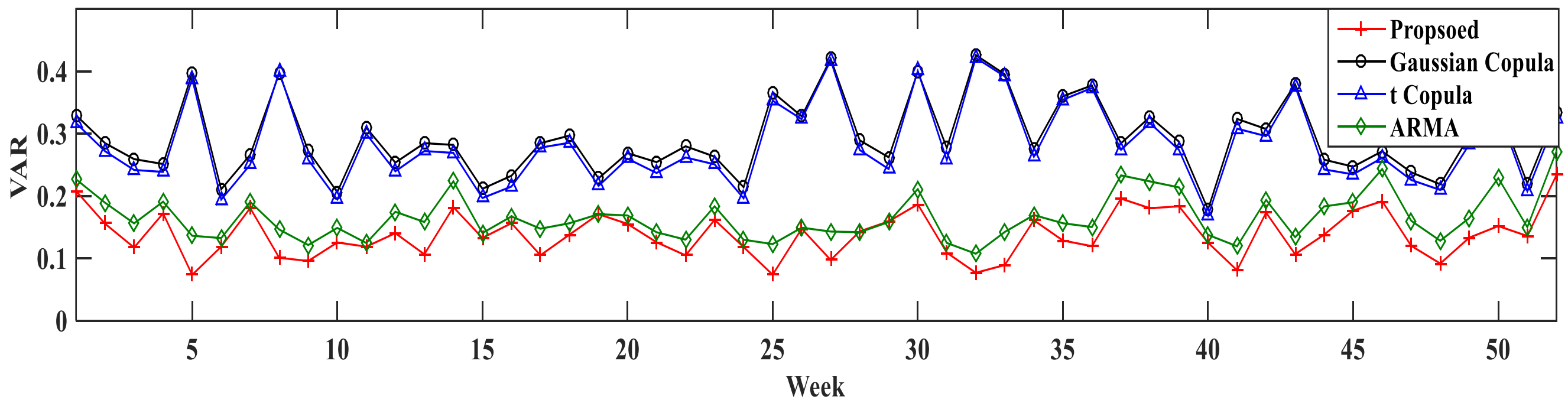

**Fig. 12** VARs of different methods in 52 weeks.

$$\text{VAR}=\frac{1}{168}\sum_{t=1}^{168}\left|\sum_{k=1}^{K}\pi_k\left(z_k^t+y^t-x^t\right)^2\right| \quad (28)$$

where $t$ stands for period. There are 168 periods in a week. $z_k$ is the forecast error in scenario $k$. $\pi_k$ is the probability of scenario $k$. $x$ is the measured wind power. $y$ is the forecast value. Both MAE and VAR are negatively oriented quantification criteria: the smaller the better.

**Table 3**
Performance of the one-year test

| Methods | MAE | VAR |
|---|---|---|
| t Copula method | 0.5168 | 0.2982 |
| Gaussian Copula method | 0.5087 | 0.2814 |
| ARMA [35] | 0.4121 | 0.1654 |
| The proposed method | 0.3542 | 0.1379 |

The test results of MAEs and VARs in every week are shown in Figs. 11 and 12. Overall, the MAE and VAR of the proposed method are smaller than those of the other three methods, indicating the scenarios generated by the proposed method can better represent real uncertainties.

Further, this paper compute MAEs and VARs for the whole year 2005. The test results are listed in Table 3. In terms of the MAE, the proposed method has 31%, 30%, and 14% decrements with respect to the t Copula, Gaussian Copula, and ARMA method, respectively. As far as the VAR is concerned, the proposed method has 53%, 51%, and 17% decrements. The rest results further validate that the proposed method outperforms other methods in a long-time test.

## 7. Conclusion

This paper develops a method that can characterize non-Gaussian and interdependent wind power forecast errors under different forecast conditions. The conditional distribution is modeled as a GMM, enhancing the scenario generation.

The proposed method has one limitation. In order to model $[\boldsymbol{X}^{\text{T}}\ \boldsymbol{Y}^{\text{T}}]^{\text{T}}$ accurately, the number of Gaussian components of a GMM may be large. It is necessary to find a way to reduce the component number without undermining the accuracy. In [29], [30], two possible solutions are discussed.

For future applications, the proposed methodology could be used to generate scenarios for various kinds of uncertain analyses, such as stochastic optimizations in the day-ahead scheduling and Monte-Carlo-based reliability analysis.

## Acknowledgements

This work was supported in part by the Foundation for Innovative Research Groups of National Natural Science Foundation of China under Grant 51621065, and in part by the Special Fund of the National Basic Research Program (973) of China under Grant 2013CB228201.

The authors would like to thank Dr. Chen-Ching Liu for his valuable discussions and suggestions. Zhiwen Wang would like to thank China Scholarship Council for supporting his study at Washington State University.

## Appendix A: Proof of the sampling method

The proof for a bivariate case is provided in the following. Those derivations are applicable to a multivariate case without major differences. As far as the $m$th Gaussian component of a GMM is concerned, there are $\omega_m C_{\text{total}}$ samples, denoted as $\boldsymbol{S}_z^m$, generated from the $m$th multivariate Gaussian distribution $N_m(\cdot)$. Without loss of generality, for arbitrary given values $z_1$ and $z_2$, it is assumed that there are $C_m$ samples, out of $\omega_m C_{\text{total}}$, falling into the area $[-\infty\ z_1]\times[-\infty\ z_2]$. An illustration is shown in Fig. A1.

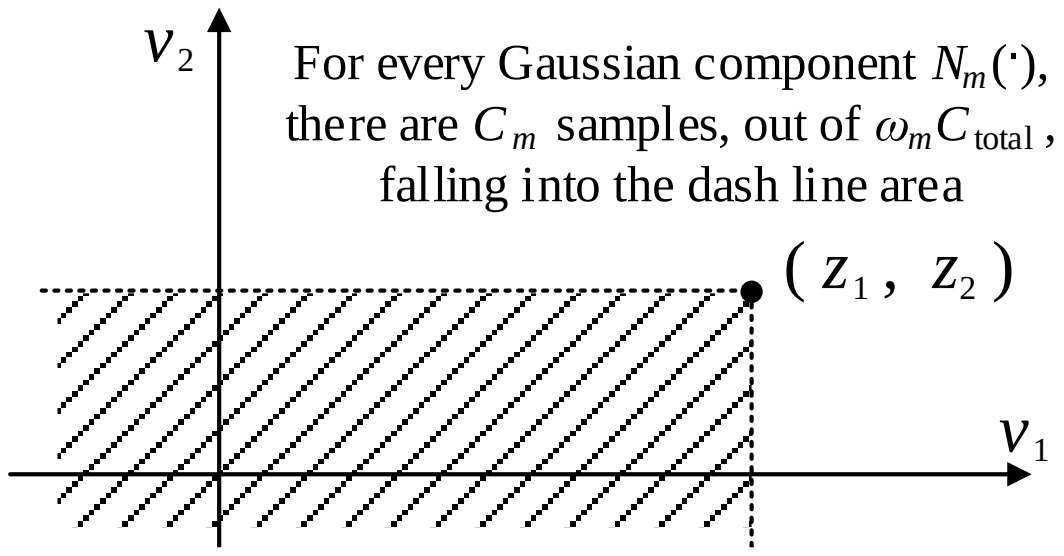

**Fig. A1** $C_m$ samples, out of $\omega_m C_{\text{total}}$, fall into the area $[-\infty\ z_1]\times[-\infty\ z_2]$

Since those samples are generated from the $m$th multivariate Gaussian distribution $N_m(\cdot)$, the following equation holds:

$$\lim_{C_{\text{total}}\to\infty}\frac{C_m}{\omega_m C_{\text{total}}}=\iint\limits_{\substack{v_1\le z_1\\ v_2\le z_2}} N_m\left(v_1,v_2\right)dv_1dv_2 \quad (A1)$$

Consider a new sample set $\boldsymbol{S}_z^{\text{GMM}}$, consisting of all samples $\boldsymbol{S}_x^m$ generated from each Gaussian component. The total number of samples of $\boldsymbol{S}_z^{\text{GMM}}$ is $C_{\text{total}}$. The number of the samples of $\boldsymbol{S}_z^{\text{GMM}}$ falling into the area $[-\infty\ z_1]\times[-\infty\ z_2]$, denoted as $C_{\text{GMM}}$, is the summation of all $C_m$. In other words, Eq. (A2) holds:

$$\lim_{C_{\text{total}}\to\infty}\frac{C_{\text{GMM}}}{C_{\text{total}}}=\lim_{C_{\text{total}}\to\infty}\frac{\sum_{m=1}^{M}C_m}{C_{\text{total}}}=\sum_{m=1}^{M}\lim_{C_{\text{total}}\to\infty}\frac{C_m}{C_{\text{total}}} \quad (A2)$$

Substituting (A1) into (A2) yields:

$$\begin{aligned}\lim_{C_{\text{total}}\to\infty}\frac{C_{\text{GMM}}}{C_{\text{total}}}&=\sum_{m=1}^{M}\left[\omega_m\iint\limits_{\substack{v_1\le z_1\\ v_2\le z_2}} N_m\left(v_1,v_2\right)dv_1dv_2\right]\\&=\sum_{m=1}^{M}\left[\iint\limits_{\substack{u_1\le z_1\\ u_2\le z_2}}\omega_m N_m\left(v_1,v_2\right)dv_1dv_2\right]\\&=\iint\limits_{\substack{v_1\le z_1\\ v_2\le z_2}}\left[\sum_{m=1}^{M}\omega_m N_m\left(v_1,v_2\right)\right]dv_1dv_2\end{aligned} \quad (A3)$$

Equation (A3) indicates that the new sample set $\boldsymbol{S}_z^{\text{GMM}}$ follows the GMM $\Sigma\omega_m N_m(\cdot)$.

## Appendix B: Different lead times

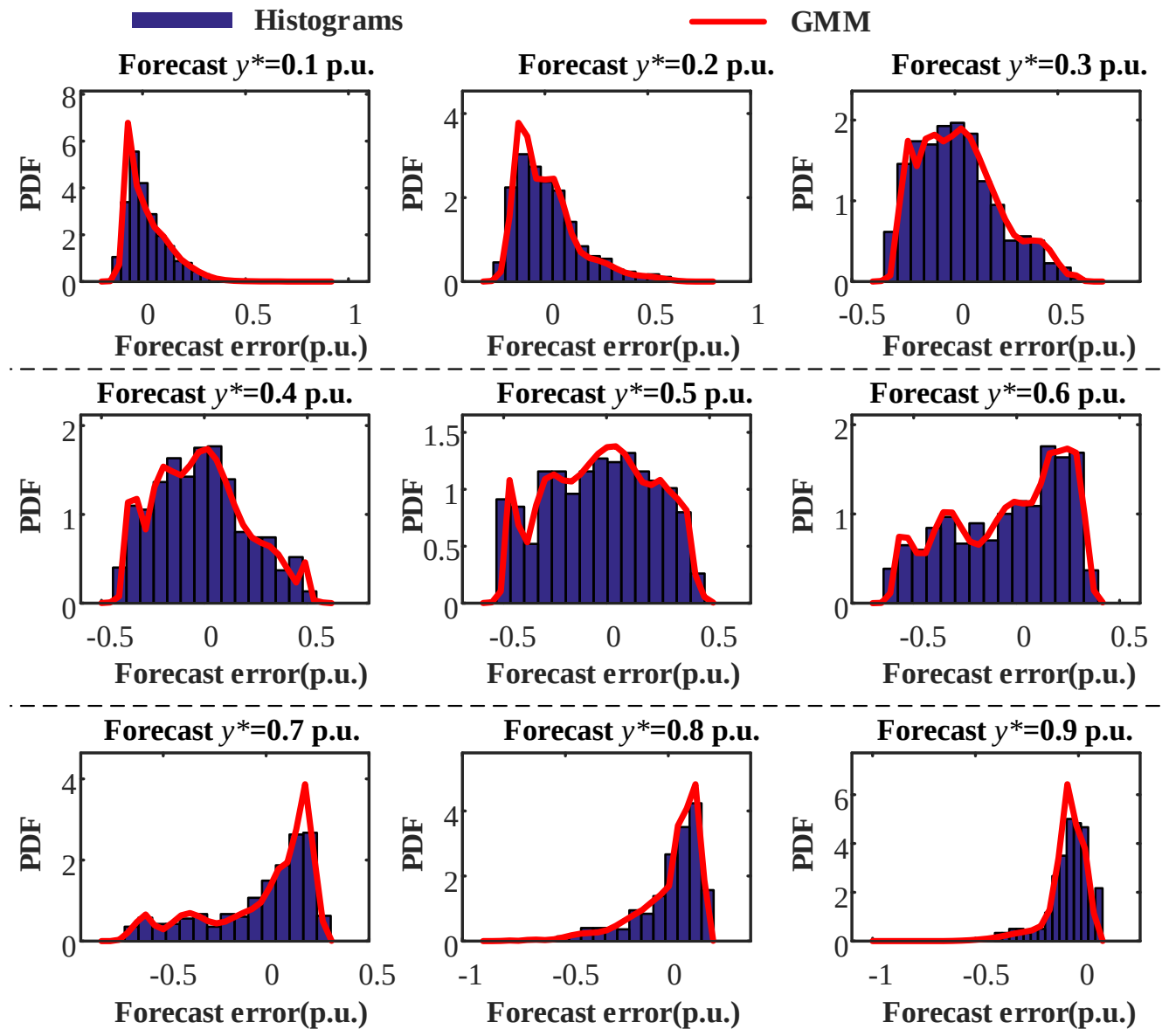

**Fig. B1** Histograms of forecast errors of the 9 bins. The forecast lead time is 6 hour. The data is from a single wind farm (ID: 4209) in the NREL "eastern wind integration data set". The data is normalized to the installed capacity.

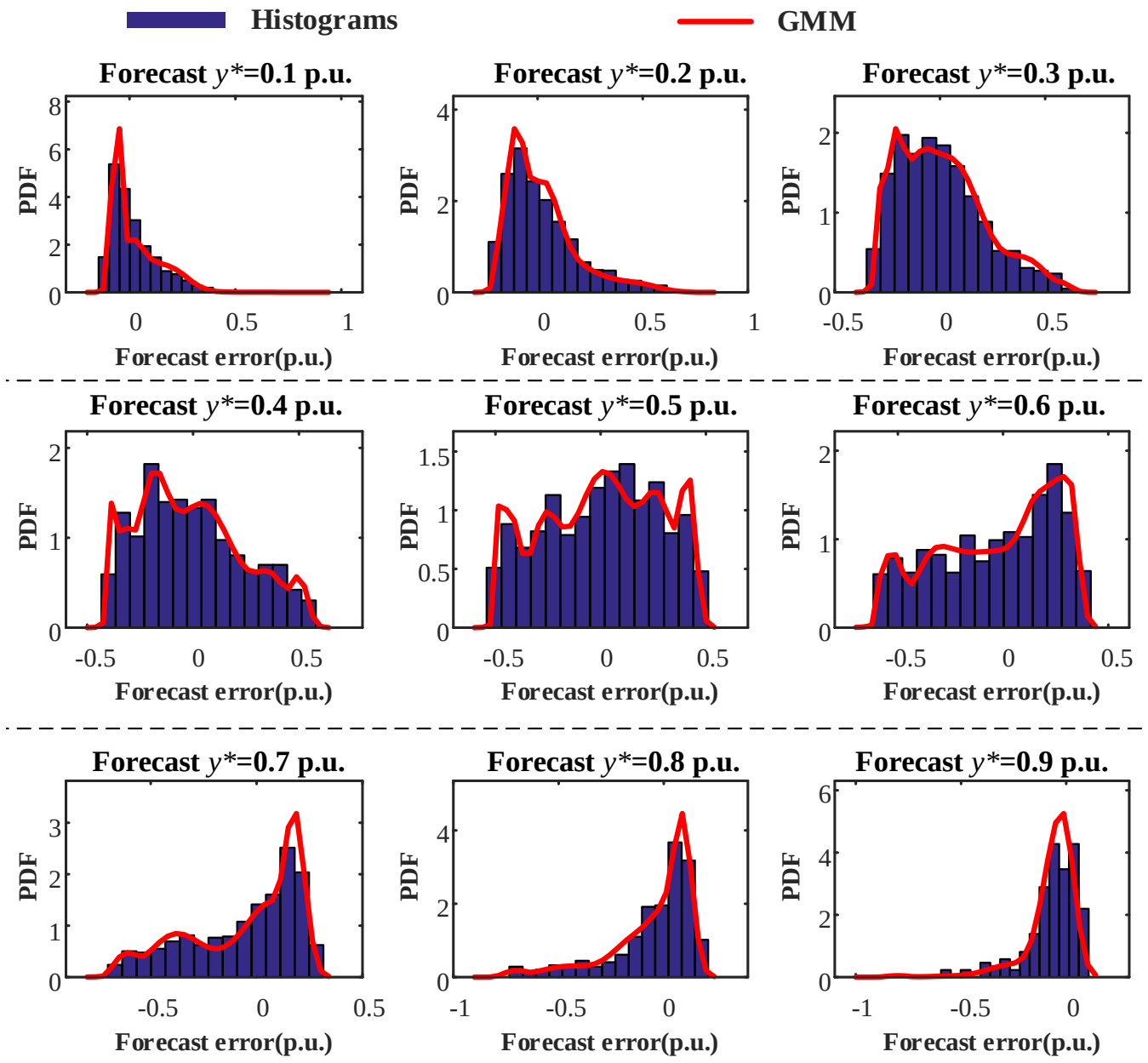

**Fig. B2** Histograms of forecast errors of the 9 bins. The forecast lead time is 24 hour. The data is from a single wind farm (ID: 4209) in the NREL "eastern wind integration data set". The data is normalized to the installed capacity.

## Appendix C: Different wind farm locations

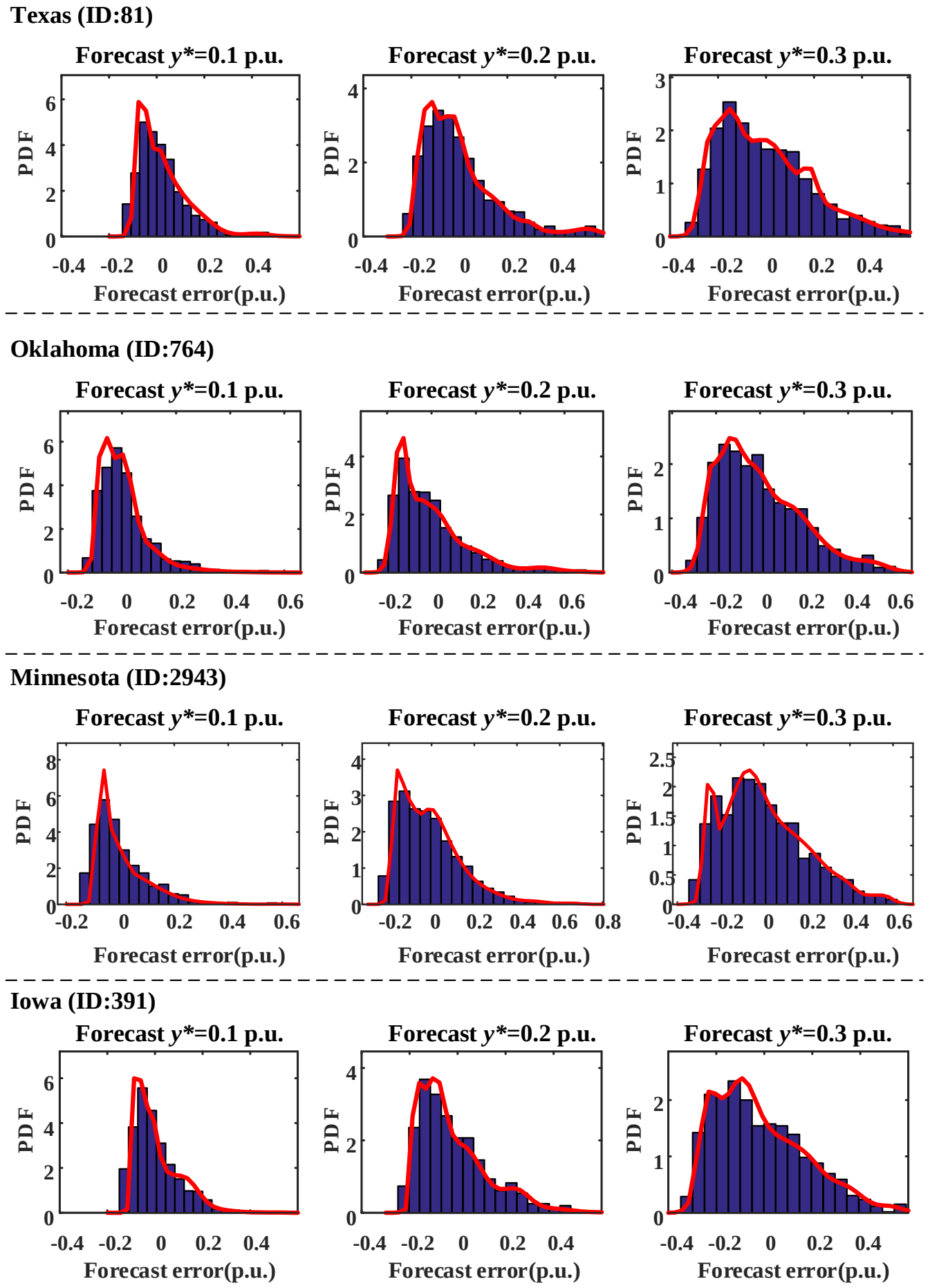

**Fig. C1** Histograms of forecast errors of the 1st, 2nd, and 3rd bins. Others are omitted. The forecast lead time is 4 hour. The data is normalized to the installed capacity.

## Appendix D: Joint distributions of forecasts and forecast errors with different lead times

In this test, the data is from a single wind farm (ID: 4209). The data is normalized to the installed capacity.

*(1) Joint distributions of forecasts with different lead times*

The joint distribution of forecasts $\boldsymbol{Y}=[\boldsymbol{Y}_{t\text{-}24}^{\mathrm{T}}, \boldsymbol{Y}_{t\text{-}6}^{\mathrm{T}}\ \boldsymbol{Y}_{t\text{-}4}^{\mathrm{T}}]^{\mathrm{T}}$ with different lead times is obtained using Eq. (17). To demonstrate the effectiveness of the proposed modeling method, this paper compares the cumulative distribution function of Eq. (17) with the empirical cumulative distribution of the historical data.

The cumulative distribution function of Eq. (17) is obtained as follows:

$$F_{\boldsymbol{Y}}(\boldsymbol{y}) = \sum_{m=1}^{M}\omega_m\left[\int\cdots\int_{\boldsymbol{v}\le\boldsymbol{y}} N_m\left(\boldsymbol{v}\ ;\ \boldsymbol{\mu}_m^{yy},\boldsymbol{\sigma}_m^{yy}\right)d\boldsymbol{v}\right] \quad \text{(D1)}$$

where $\boldsymbol{v}$ is the integral variable; $\boldsymbol{y}$ is the upper limit. The multiple integral in Eq. (D1) can be computed using the *mvncdf* function in MATLAB.

The empirical cumulative distribution of the historical data is obtained as follows:

$$\tilde{F}_{Y}(y) = \frac{1}{n}\sum_{i=1}^{n} I_i(\tilde{y}_i, y) \tag{D2}$$

$$I_i(\tilde{y}_i, y) = \begin{cases} 1 & \text{if} \quad \tilde{y}_i \le y \\ 0 & \text{else} \end{cases} \tag{D3}$$

where $n$ is the total number of the historical data records; $\tilde{y}_i$ denotes the $i$th data record. $I_i(\tilde{y}_i, y)$ is an indicator function: if every entry of $\tilde{y}_i$ is smaller than that of $y$, $I_i(\tilde{y}_i, y)$ is 1; otherwise, $I_i(\tilde{y}_i, y)$ is 0.

This paper compares $F_Y(y)$ and $\tilde{F}_Y(y)$ in 1000 different values of $y$. The test results are shown in Fig. D1. It can be observed that the GMM-based cumulative distribution matches well with the empirical cumulative distribution. The maximum error is 0.0087. Such results indicate that the joint distribution in Eq. (17) is a good representation for the wind power forecasts.

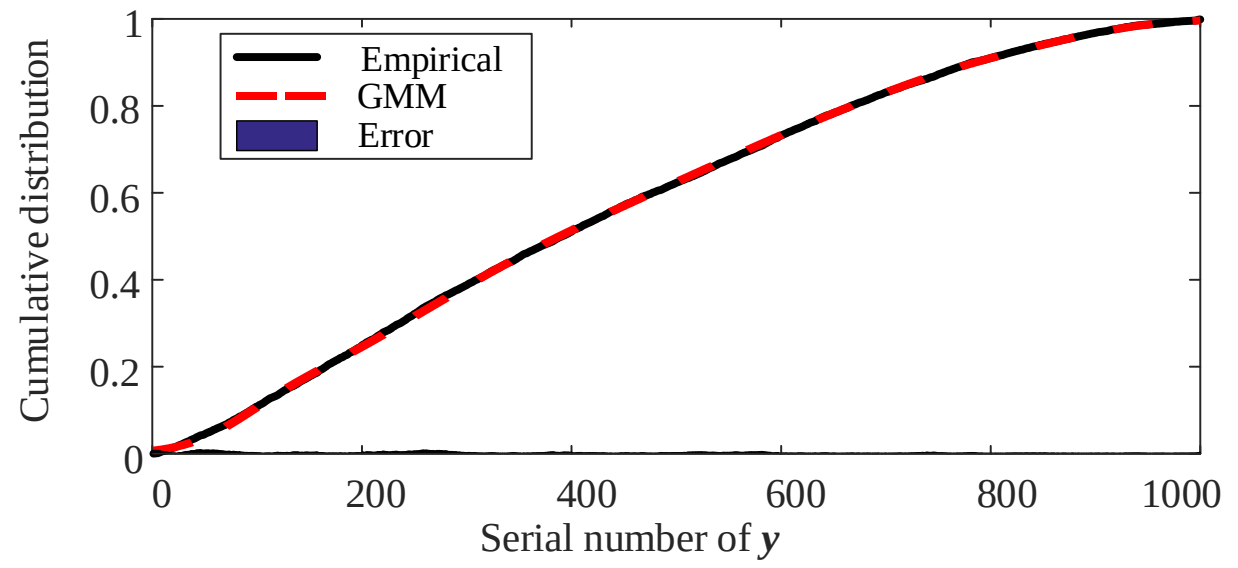

**Fig. D1** Cumulative distributions of forecasts in 1000 different $y$.

*(2) Joint distributions of forecast errors with different lead times*

The joint distributions of forecast errors with different lead times are obtained using Eqs. (18)-(20). The obtained distribution is also compared with the empirical cumulative distribution of the historical data of forecast errors. The test results are shown in Fig. D2.

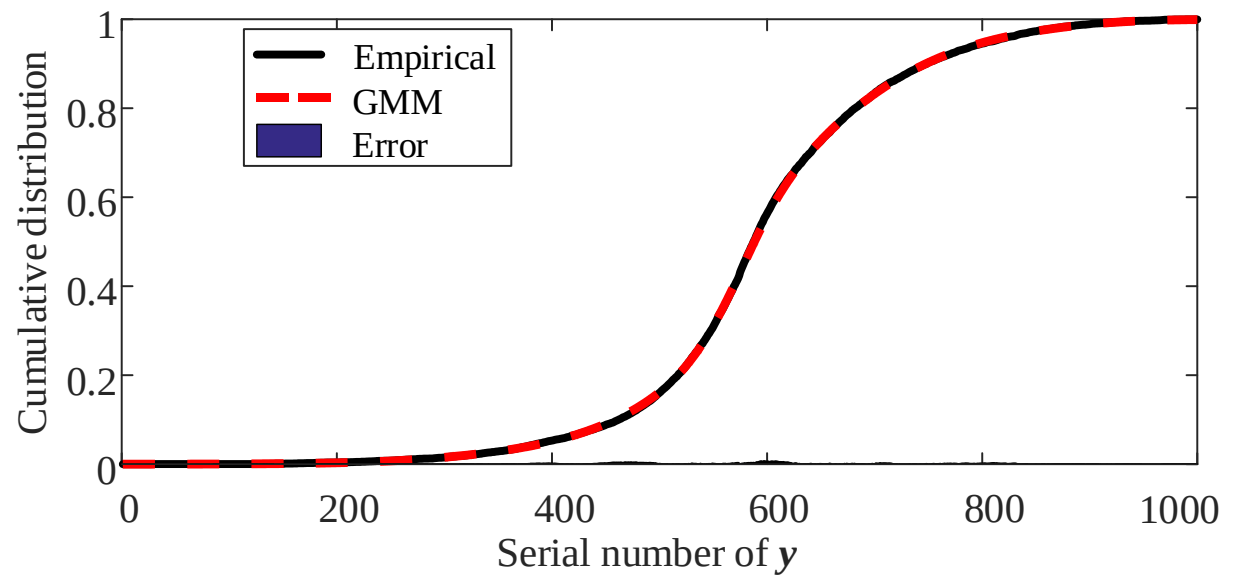

**Fig. D2** Cumulative distributions of forecast errors in 1000 different $y$.

It can be seen that the GMM-based joint distribution of forecast errors coincides with the historical data, indicating this joint distribution can well represent the uncertainty of forecast errors with different lead times.